\documentclass[%
aip,
rsi,
 amsmath,amssymb,
 reprint,%
]{revtex4-1}

\usepackage{graphicx}% Include figure files
\usepackage{dcolumn}% Align table columns on decimal point
\usepackage{bm}% bold math
\usepackage[utf8]{inputenc}
\usepackage[T1]{fontenc}
\usepackage{mathptmx}
\usepackage{etoolbox}
\usepackage[version=4]{mhchem}
\usepackage[percent]{overpic}
\usepackage[usestackEOL]{stackengine}
\usepackage{siunitx}
\usepackage{color}
\usepackage[colorlinks=true,linkcolor=black,citecolor=black,urlcolor=black]{hyperref}

\DeclareSIUnit\bar{bar}
\newcommand{\white}[1]{\textcolor{white}{#1}}

\newcommand{\figtext}[1]{\scriptsize{\makebox[0pt]{\Centerstack{#1}}}}

\makeatletter
\def\@email#1#2{%
 \endgroup
 \patchcmd{\titleblock@produce}
  {\frontmatter@RRAPformat}
  {\frontmatter@RRAPformat{\produce@RRAP{*#1\href{mailto:#2}{#2}}}\frontmatter@RRAPformat}
  {}{}
}%
\makeatother
\begin{document}

\preprint{AIP/123-QED}

\title[Coulomb Blockade Thermometry Based Nanocalorimetry]{Coulomb Blockade Thermometry Based Nanocalorimetry}
% Force line breaks with \\
\author{Mari C. Cole}
\altaffiliation{M. C. Cole, M. T. Pelly and C. V. Topping  contributed equally to this work}
\author{Maximilian T. Pelly}
\altaffiliation{M. C. Cole, M. T. Pelly and C. V. Topping  contributed equally to this work}
\author{Craig V. Topping}
\altaffiliation{M. C. Cole, M. T. Pelly and C. V. Topping  contributed equally to this work}
\affiliation{SUPA, School of Physics and Astronomy, University of St Andrews, North Haugh, St. Andrews, Fife KY16 9SS, United Kingdom}

\author{Thomas Reindl}
\author{Ulrike Waizmann}
\author{J\"{u}rgen Weis}
\email{j.weis@fkf.mpg.de}
\affiliation{Max Planck Institute for Solid State Research, Heisenbergstr. 1, D-70569 Stuttgart, Germany}

\author{Andreas W. Rost}
\email{a.rost@st-andrews.ac.uk}

\affiliation{SUPA, School of Physics and Astronomy, University of St Andrews, North Haugh, St. Andrews, Fife KY16 9SS, United Kingdom}

\date{}

\begin{abstract}
Specific heat is a powerful probe offering insights into the entropy and excitation spectrum of the studied material. While it is well established, a key challenge remains the measurements of microcrystals or thin films especially in the sub-Kelvin, high magnetic field regime. Here we present a setup combining the high sensitivity of \ce{SiN_x} membrane based calorimetry with the absolute accuracy of Coulomb blockade thermometry to realise a nanocalorimeter for such tasks. The magnetic field independent technique of Coulomb blockade thermometry provides an on-platform thermometer combining a primary thermometry mode for in-situ calibration with a fast secondary mode suitable for specific heat measurements. The setup is validated using measurements of a 20~\si{\ug} sample of \ce{Sr3Ru2O7} achieving a resolution on the order of 0.1~\si{\nano\joule\per\kelvin} at 500~\si{\milli\kelvin} and an absolute accuracy limited by the determination of the sample's mass. Measurements of \ce{CeRh2As2} further highlight the benefits of measuring microcrystals with such a device.
\end{abstract}

\maketitle

\section{\label{sec:Introduction}Introduction}

The study of new quantum materials is of great interest, both as a way to probe fundamental questions and with the aim of developing technological applications.\cite{Becher2023} Materials research is often driven by the search for and discovery of new phases and phase transitions. The evolution of entropy across these contains important information on the excitation spectrum of the underlying ground states.\cite{Ashcroft1976}
Calorimetry is therefore a particularly important technique to study quantum materials, giving direct insight into entropy and thus density-of-state changes, and allowing straightforward tracking of electronic phase transitions. While the technique is generally well established,\cite{Stewart1983} conventional setups often have a significant addenda heat capacity limiting their suitability to measurements of samples of \si{\milli\gram} mass or larger. However there are important material classes where single crystals of this size are unavailable. These include for example new materials, where sample growth is not optimized for large single crystals, such as the \ce{La}-Nickelates;\cite{Li2019, Krieger2022, Catalano2018} established materials where large high quality single crystals have been effectively impossible to grow, such as Tl2201;\cite{Peets2010} and thin film heterostructure materials where large sample masses are impossible by definition.\cite{Bahri2017, Takachi2013a, Takachi2013b, Matsuda2013} 
In recent years a number of nanocalorimeters have been developed advancing our capabilities in measuring \si{\ug}-samples with sub \si{nJ/K} heat capacities.\cite{Yi2019}
Of those, devices based on mm-scale silicon nitride (\ce{SiN_x}) membranes with sub-200~\si{\nm} thickness show high promise for low temperature operation due to their low thermal conductivities and negligible addenda.\cite{Fon2005, Cooke2008, Tagliati2012, Wickey2015, Willa2017}

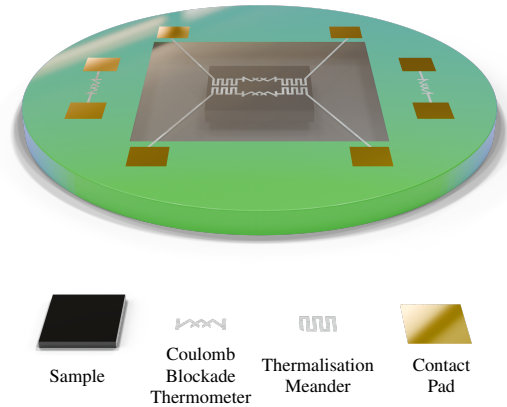
\begin{figure}[t]
    \centering
    \begin{overpic}[width=\linewidth]{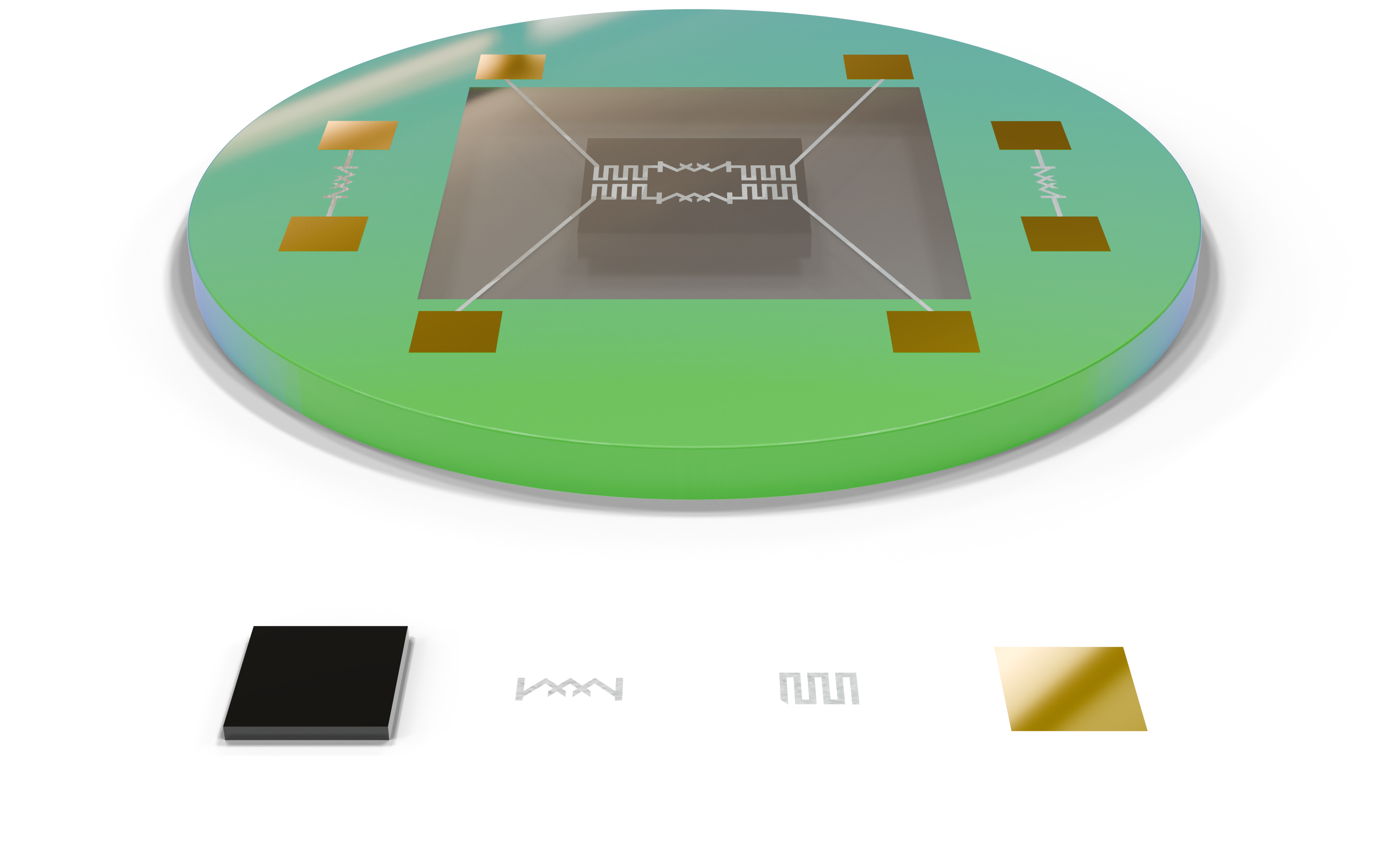}
    \put (22,4) {\figtext{Sample}}
    \put (41,4) {\figtext{Coulomb\\Blockade\\Thermometer}}
    \put (59,4) {\figtext{Thermalisation\\Meander}}
    \put (78, 4) {\figtext{Contact\\Pad}}
    \end{overpic}
    \caption{A not-to-scale schematic representation of a completed nanocalorimeter.
    The figure shows a representation of high-temperature Coulomb blockade thermometers with thermalisation meanders (silver) deposited on top of a silicon nitride membrane.
    The sample to be measured (black) is mounted on the other side of the membrane.
    Gold contact pads deposited on the silicon chip frame (green) allow interfacing with the device.}
    \label{fig:setup schematic}
\end{figure}

However for measurements in the important sub-Kelvin, high magnetic field regime high accuracy remain a challenge, with one source of uncertainty being secondary resistive thermometry.\cite{Tagliati2012, Cryogenic2024, Fortune2023, PalmerFortune2024}
Such thermometry requires extensive calibration, especially if used in a magnetic field, with residual uncertainty due to changes in resistance by repeated thermal cycling of membranes.\cite{Balle2000}
Calibration can also become effectively impossible at lowest temperatures as calibrating a resistive thermometer requires it to be in thermal equilibrium with a known bath temperature.
This can be challenging to achieve in conventional cryogenic systems without the installation of complex heat switches\cite{Roshanzadeh2020} as the usual approach of ex-situ calibration is unavailable for membrane based devices.\cite{Rost2009b, Tagliati2012} 

The solution pursued by us (see Fig.~\ref{fig:setup schematic}) is the inclusion of a primary Coulomb blockade thermometer on the \ce{SiN_x} membrane.
Coulomb blockade thermometry is a well established thin film based thermometry technology that has a slow primary and fast secondary measurement mode, and in addition is magnetic field independent.\cite{Pekola1994, Farhangfar1997, Kauppinen1998, Farhangfar2000, Farhangfar2001, Pekola2002, Feshchenko2013, Meschke2016, Sarsby2020} As such, Coulomb blockade thermometers (CBTs) are perfect for our device, combining negligible heat capacity and thermal conductivity inherent in thin film technology with the advantages of magnetic field independent primary thermometry. The measurement range is determined by the design of the CBTs and can be tuned for operation from tens of Kelvin\cite{Meschke2016} down to the \si{\micro\kelvin} regime.\cite{Sarsby2020} A schematic of our prototype device is shown in Fig.~\ref{fig:setup schematic}, highlighting the key structures. The calorimeter has been successfully operated in a Kelvinox 25 dilution refrigerator equipped with a \qty{9}{\tesla} / \qty{1}{\tesla} / \qty{1}{\tesla} vector magnet.
Using samples with a mass of approximately 20~\si{\ug} (typical side length less than 100~\si{\um}), we have resolved heat capacity with a resolution well below 0.1~\si{\nano\joule\per\kelvin} and an absolute accuracy limited by the knowledge of the sample mass.

An introduction to CBTs and their various relevant operating modes is given in section~\ref{sec:CBT background}. In section~\ref{sec:device} we describe the design and fabrication of our device.
The design parameters were guided by numeric simulations discussed in section~\ref{sec:simulation}, which also allowed prediction of characteristic timescales and the significance of geometric correction factors due to sample dimensions and placement.
Validation of the setup using the well studied material \ce{Sr3Ru2O7}\cite{GRIGERA2001, GRIGERA2004, Borzi2007, Rost2009a, Rost2009b, Mercure2010, Rost2010, Rost2011, Bruin2013, Sun2018, Lester2015} is covered in section~\ref{sec:validation}. In section~\ref{sec:CeRh2As2} we present a study of the topical material \ce{CeRh2As2}\cite{Khim2021, Hafner2022, Semeniuk2023} to highlight the benefits of being able to study microcrystals.

\section{\label{sec:CBT background}Coulomb Blockade Thermometry}
\begin{figure}[t!]
    \includegraphics[width=1\linewidth]{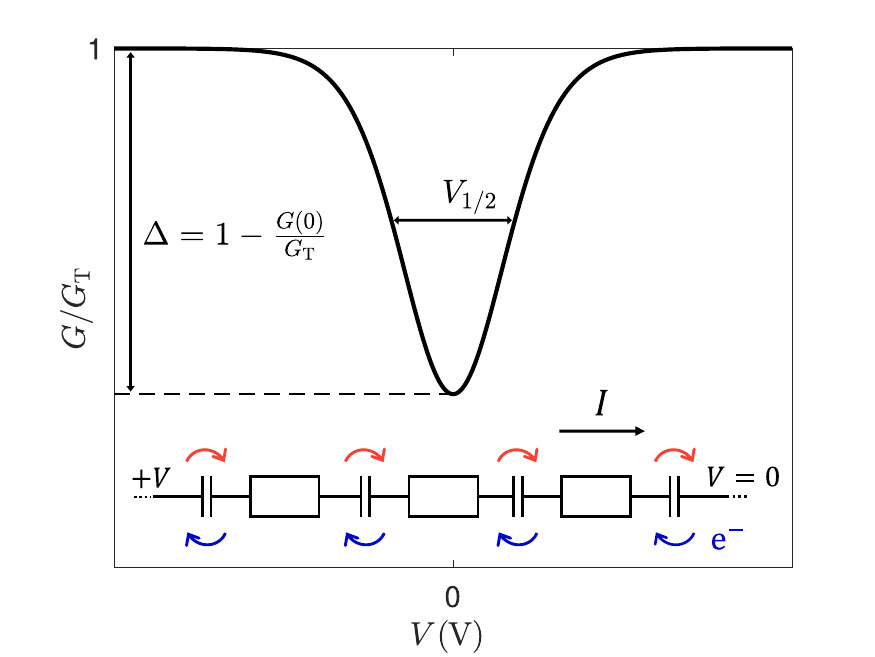}
    \caption{An example of the normalized differential conductance G through a CBT array in the weak CB regime. The key features for thermometry are the full width at half minimum \(V_{1/2}\) and the conductance dip depth \(\Delta\). Inset: A schematic of a single CBT array.}
    \label{fig:dip}
\end{figure}

\subparagraph{High-temperature regime}
At the core of our calorimeter is the addition of Coulomb blockade thermometers (CBTs) to a \ce{SiN_x} membrane based setup. The use of the Coulomb blockade effect for thermometry was first developed by Pekola \textit{et al.} \cite{Pekola1994} building on work by Averin and Likarev.\cite{Averin1986} CBTs are constructed out of arrays of tunnel junctions  connecting small metallic islands (see insert in Fig.~\ref{fig:dip}).\cite{Pekola1994} These islands are characterized by a single-electron  charging energy $E_C$ and the junctions by an effective resistance $R_J$. In order for a current to flow, continuous tunneling of electrons is required. At low bias voltages $V<<E_C/e$ the charging energy $E_C$, which is dominated by the capacitance due to island overlap at the junctions,  can prevent individual tunneling events and therefore current flow. At very low temperature $k_BT<<E_C$ in the well known strong Coulomb blockade regime this leads to insulating behavior\cite{Averin1986} while at higher temperatures thermal fluctuations induce tunneling.\cite{Pekola1994} This high-temperature behavior results in a characteristic  temperature dependent dip structure in the differential conductance \(G = \partial I / \partial V\) (see Fig.~\ref{fig:dip}).

For an array of $N$ tunnel junctions the effective charging energy $E_C$ is given by
\begin{equation}
    E_C = \frac{N-1}{N} \frac{e^2}{C} \; ,
\end{equation}
where \(e\) is the elementary charge and \(C\) the capacitance of a single island. Pekola \textit{et al.} \cite{Pekola1994} derived a master equation for excess charge on the islands considering individual tunneling events governed by Fermi's golden rule. They showed that in the limit of high-temperature this can be solved exactly resulting in a differential conductance of the form 
\begin{equation}
\frac{G\left(T, V\right)}{G_T} = 1 - \frac{E_C}{k_BT}\;g\left(\frac{eV}{Nk_BT}\right)\;.
\label{eqn:dip equation}
\end{equation}
Here \(G_T\) is the infinite bias conductance and \(g(\nu)\) is a function given by

\begin{equation}
    g(\nu) = \frac{\nu\sinh{(\nu)}-4\sinh^2{(\nu/2)}}{8\sinh^4{(\nu/2)}}\;.
    \label{eqn:dip g function}
\end{equation}

This differential conductance response~\eqref{eqn:dip equation} is independent of magnetic field.\cite{Pekola2002, Meschke2016}  In Fig.~\ref{fig:dip} we show an example conductance curve highlighting the two key temperature dependent features relevant for thermometry: the full width at half minimum (FWHM) \(V_{1/2}\) and the relative conductance dip depth \(\Delta\). 

\begin{figure}
    \centering
    \includegraphics[width=\linewidth]{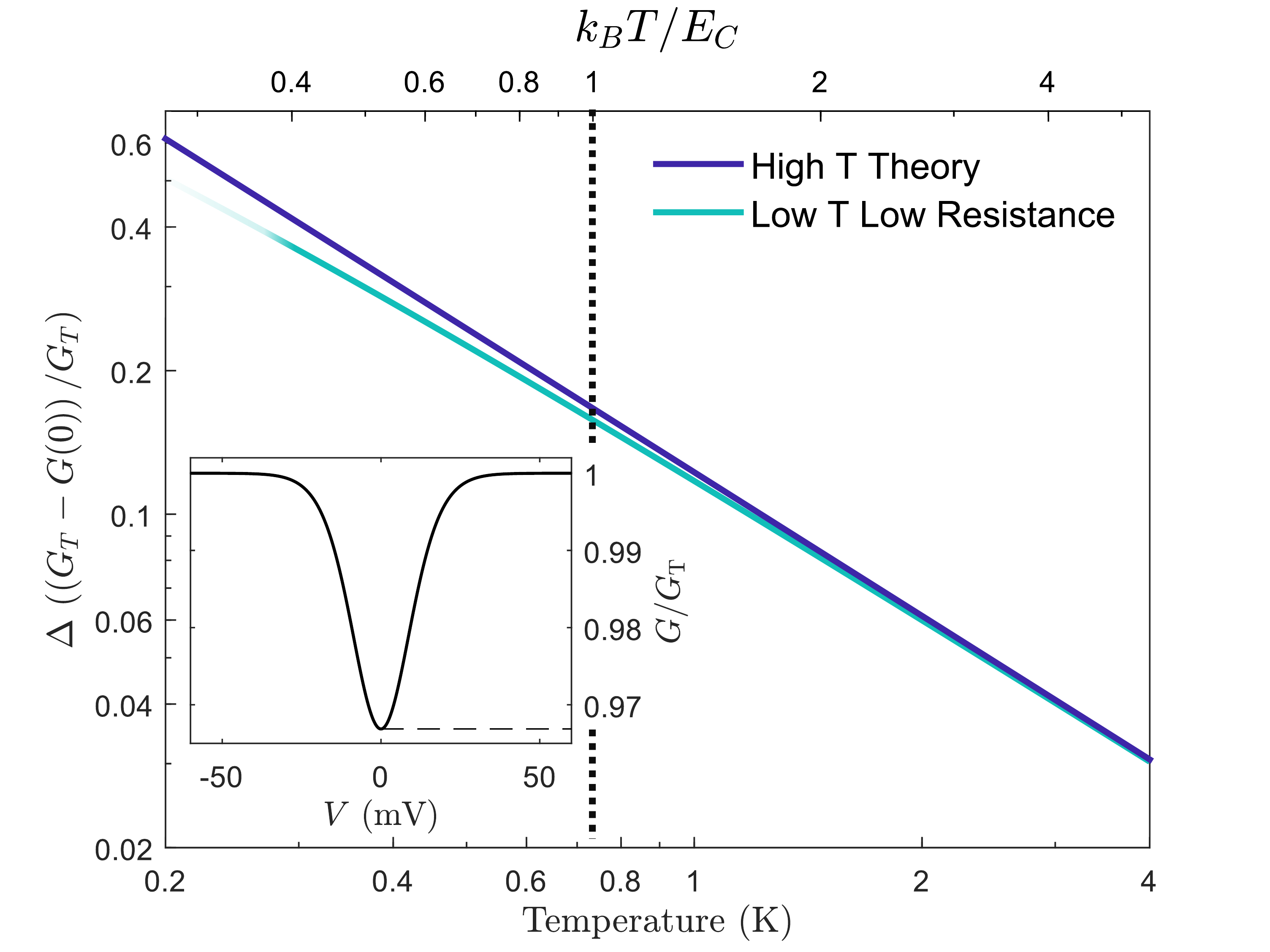}
    \caption{Theoretical temperature dependencies of \(\Delta\) for the high and low $T$ theories. The CBT parameters are comparable to those used in later sections with an $E_C$ of  \qty{63.3}{\micro\electronvolt} and $G_T$ of \qty{18.8}{\micro\ampere\per\volt}. This gives a corresponding junction resistance of \qty{13.3}{\kilo\ohm} with cutoff temperatures \(E_C/k_B = 735\)~\si{\milli\kelvin} (high-temperature theory, dashed line) and \(0.4E_C/k_B = 294\)~\si{\milli\kelvin} (loss of universality, fade out). Inset: An example of a \(G(V)\) trace at \qty{4}{\kelvin} for the same device.}
    \label{fig:G0 comparison}
\end{figure}

The FWHM provides a primary thermometery mode, as its proportionality to temperature depends only on constants:\cite{Pekola1994}
\begin{equation}
\label{eqn:CBT T HTHR FWHM}
    T = \frac{V_{1/2}e}{5.439Nk_B} \; .
\end{equation}
$\Delta$ on the other hand provides a secondary thermometry mode as it is related to $T$  via $E_C$:
\begin{equation}
\label{eqn:CBT T HTHR DGOGT}
    T = \frac{E_C}{6k_B\Delta} \; ,
\end{equation}
where \(\Delta = 1 - G(0)/G_T\) and \(G(0)\) denotes the differential conductance at zero bias (the bottom of the dip in Fig.~\ref{fig:dip}). For secondary thermometry mode measurements, $E_C$ can be determined at a single stable temperature point via the primary thermometry mode. We choose to do so at the temperature of our liquid helium bath as flooding the inner vacuum chamber with exchange gas ensures a common, stable temperature of all thermometers in the setup. The secondary mode thereafter simply requires the measurement of the zero bias differential conductance \(G(0)\)\cite{Pekola1994} which we demonstrate to be sufficiently fast for calorimetry measurements.

\subparagraph{Low-temperature regime}The above theoretical framework has been extended to temperatures well below $T=E_C/k_B$ for junction resistances $R_J$ both large and small on the scale of the von Klitzing constant $R_K=h/e^2$ $\approx$ 25~\si{\kilo\ohm}.\cite{Farhangfar1997, Farhangfar2000, Farhangfar2001, Feshchenko2013}
For resistances larger than $R_K$ Farhangfar \textit{et al.} derived an expansion of $\Delta$ as a function of $E_C \sim k_B T$:\cite{Farhangfar1997}
\begin{equation}
	\label{eqn:DG_GT_3rd}
	\Delta = \frac{1}{6} \frac{E_C}{k_B T} - \frac{1}{60} \left( \frac{E_C}{k_B T}\right)^2 + \frac{1}{630} \left( \frac{E_C}{k_B T}\right)^3 + \cdots.
\end{equation}
Feshchenko \textit{et al.} report that with the inclusion of terms up to third order, measurements at temperatures down \(k_BT/E_C \sim 0.4 \) are possible, with the low temperature limit set by the loss of universal behavior around this point.\cite{Feshchenko2013}
For low resistance junctions Farhangfar \textit{et al.} \cite{Farhangfar2000, Farhangfar2001} considered fluctuations in the current and phase across each junction finding an analytical relationship between \(\Delta\) and temperature given by 

\begin{equation}
\label{eqn:CBT T LTLR DGOGT}
    \Delta = \frac{N-1}{N}\frac{e^2R_J}{\pi\hbar}\left(\Psi{\left(1+u_R\right)}+\gamma+u_R\Psi'{\left(1+u_R\right)}\right)
\end{equation}
where \(\Psi\) is the digamma function, \(\gamma\) is Euler's constant\footnote{Equal to \(-\Psi\left(1\right)\) or approximately 0.5772156649...} and the parameter \(u_R\) is defined by
\begin{equation}
u_R = \frac{E_C}{4\pi^2 k_BT} \frac{R_K}{R_J}  \; .  
\end{equation}
This theory has been shown by Farhangfar \textit{et al.} to be valid for junction resistances of $2$ to $23$~\si{\kilo\ohm} to temperatures as low as  \(k_BT/E_C \sim 0.4\).
\cite{Farhangfar2000, Farhangfar2001}
For practical application equations~\eqref {eqn:DG_GT_3rd} and~\eqref{eqn:CBT T LTLR DGOGT} can be inverted numerically to obtain a $T(\Delta)$ relationship and in the high-temperature limit both reduce to~\eqref{eqn:CBT T HTHR DGOGT}.

In Fig.~\ref{fig:G0 comparison} we show for a typical $E_C$  and $G_T$ of one of our devices (63.3~\si{\micro\electronvolt} and 18.8~\si{\micro\ampere\per\volt} respectively, giving $R_j = 13.3$~\si{\kilo\ohm}) the extrapolated high-temperature behavior for reference (blue) together with the low junction resistance expansion (cyan).
These diverge rapidly below $T = 2E_C/k_B$, highlighting the importance of using the correct theory.

\subparagraph{Self-Heating} As measurements in primary mode require significant bias voltages to be applied, the thermometer on the \ce{SiN_x} membrane can be affected by self-heating even in exchange gas at \qty{4}{\kelvin}. Such heating can result in a change in the dip shape, narrowing it around zero bias and broadening the tails, introducing a systematic uncertainty. \cite{Kauppinen1996} This can affect CBTs on membranes but is negligible for CBTs deposited on the thermally well anchored membrane frames. In such cases one can still calibrate the charging energy $E_C$ via the temperature dependence of the zero bias conductivity which is unaffected by such self-heating. In the high-temperature limit in which  equation~\eqref{eqn:CBT T HTHR DGOGT} holds it can be seen that for two thermometer the conductance dips \(\Delta_M\) (membrane) and \(\Delta_F\) (frame) are directly proportional with the linear proportionality factor $E_{C,M}/E_{C,F}$:
\begin{equation}
    \Delta_M = \Delta_F\frac{E_{C,M}}{E_{C,F}} \; .
    \label{eqn:delta-delta}
\end{equation}
One can therefore use a \(\Delta_M\) -\(\Delta_F\) plot over a relevant temperature regime to calibrate the membrane thermometer in secondary mode against a thermally well anchored frame thermometer for which a primary mode measurement is possible. This calibration only relies on the assumption that for small excitations the membrane and frame thermometer are in good thermal contact which we have generally found to be the case down to approximately \qty{500}{\milli\kelvin}. As we show in Appendix \ref{ap:cal} a violation of that assumption would result in a deviation from linearity of the  \(\Delta_M-\Delta_F\) relationship. As shown in Appendix \ref{ap:LR cal} this procedure can be extended to the low temperature regime of the CBTs.

\section{\label{sec:device}Design and Fabrication}

\subsection{Calorimeter Design}
\begin{figure}[t]
    \centering
    \includegraphics[width=1\linewidth]{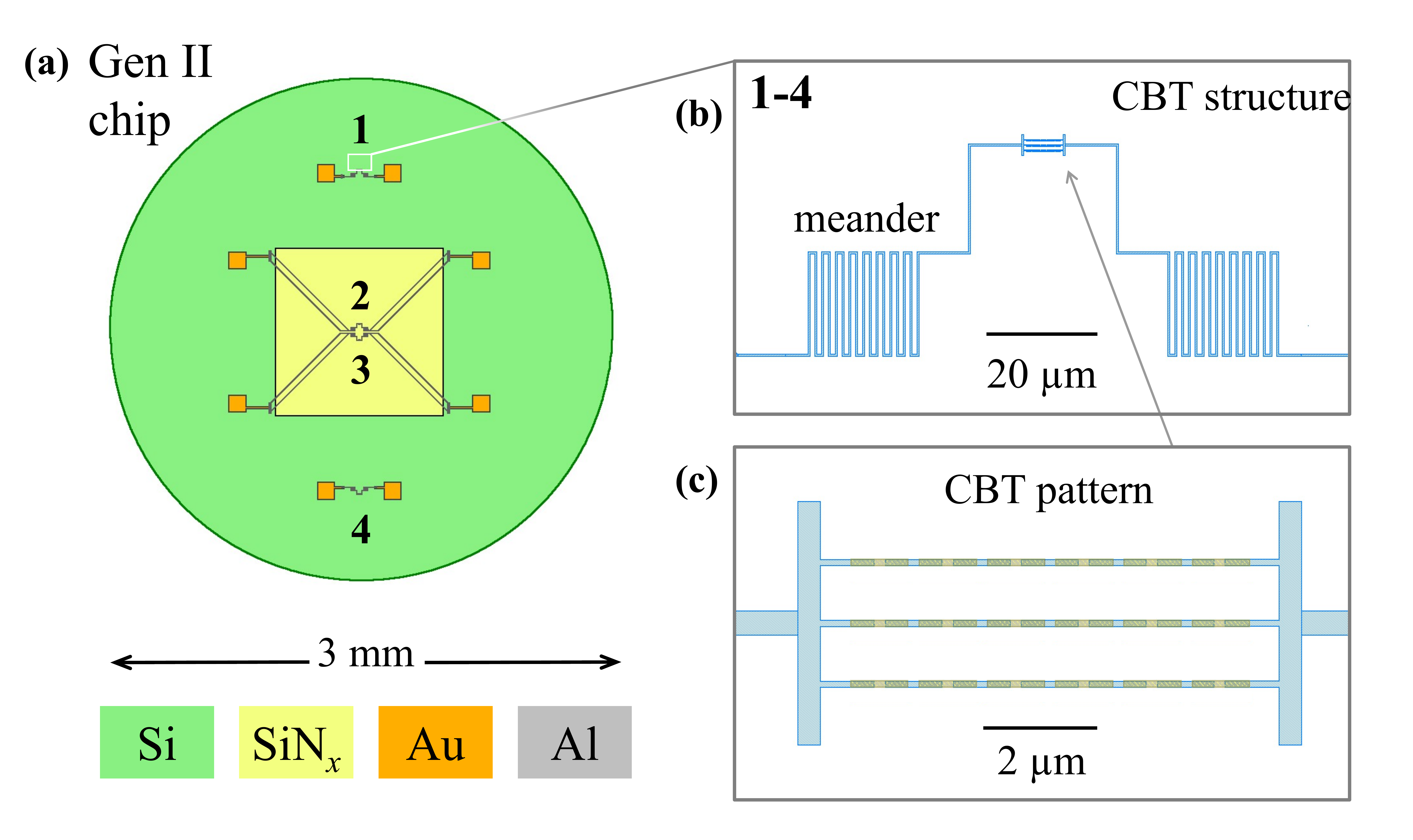}
    \caption{The CBT chip design.
    There are two CBTs on the silicon frame, and two CBTs on the silicon nitride window.
    Each CBT may be used as a heater or thermometer.
    \textbf{a}: The silicon chip with the lithography pattern for the Au contacts and Al structures and wires.
    \textbf{b}: The thermometry setup in the center of the membrane showing the CBT structure and thermalisation meanders.
    \textbf{c}: The target tunnel junction array. Different colors denote islands separated in the out of page direction by an insulating \ce{Al2O3} layer. Note orphaned islands produced by the shadow mask process are not shown here for clarity. 
    }
    \label{fig:chip}
\end{figure}

The key design criteria for the calorimeter are the arrangement of CBTs on the chip and the design of the tunnel junctions themselves; with the aim of using a minimalist layout for our proof-of-principle study. We use two CBTs located in the center of the membrane with thermalisation meanders on either side in order to ensure good contact with the sample. One of these is used as a heater and the other as a thermometer, allowing a measure of redundancy. Additional CBTs are located on the frame of the chip for calibration and as thermometers or heaters for temperature stabilization. A typical layout is shown in Fig.~\ref{fig:chip}.

\begin{figure}[b]
    \centering
    \begin{overpic}[width=0.9\linewidth]{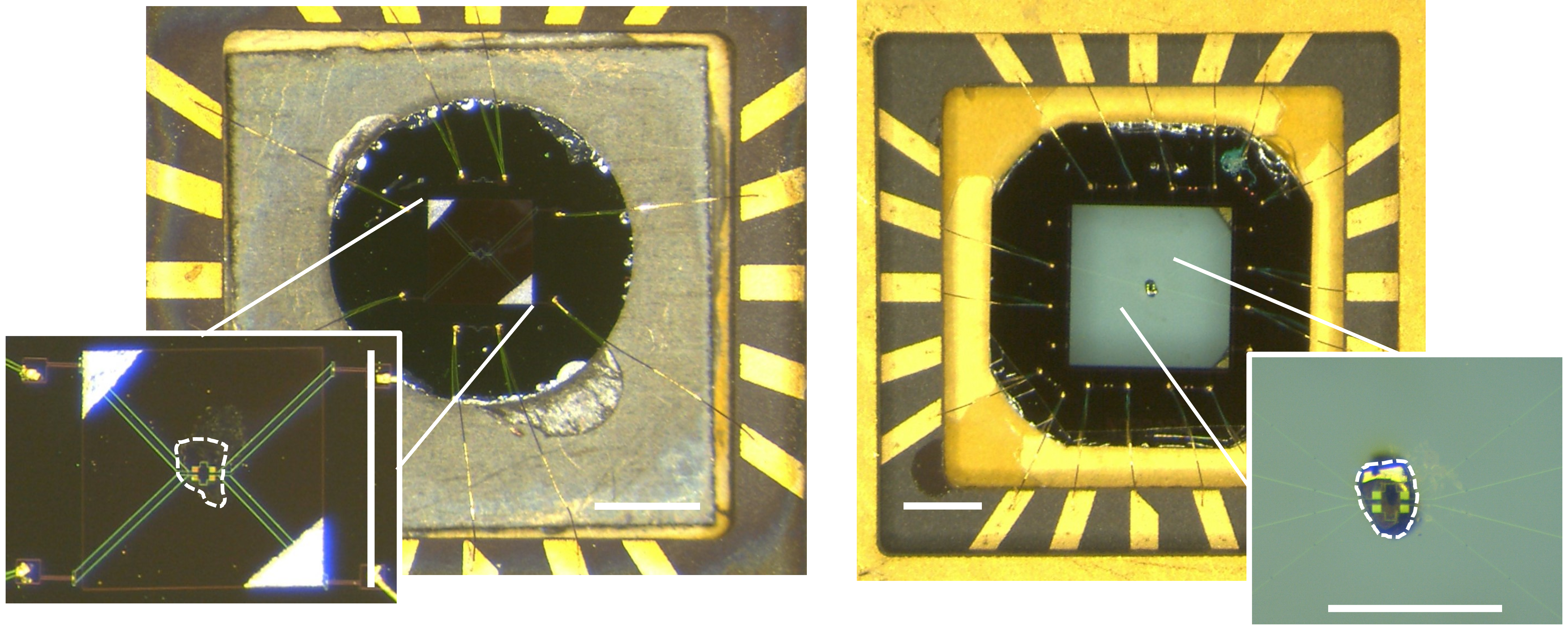}
        \put (0, 42) {(a)}
        \put (15.5, 9) {\white{\scriptsize{1~\si{\milli\meter}}}}
        \put (37.5, 9) {\white{\scriptsize{1~\si{\milli\meter}}}}
        \put (100, 42) {(b)}
        \put (57, 9) {\white{\scriptsize{1~\si{\milli\meter}}}}
        \put (85.25, 2.5) {\scriptsize{\white{0.5~\si{\milli\meter}}}}
    \end{overpic}
    \caption{Examples of some completed calorimeters.
    \textbf{a}: A small membrane device used for the \ce{Sr3Ru2O7} validation study.
    \textbf{b}: A larger membrane device used for the CeRh$_2$As$_2$ study. 
    Insets show a zoomed in view of the sample outlined in white.
    Note the different scales of the images.} 
    \label{fig:CBT pics}
\end{figure}
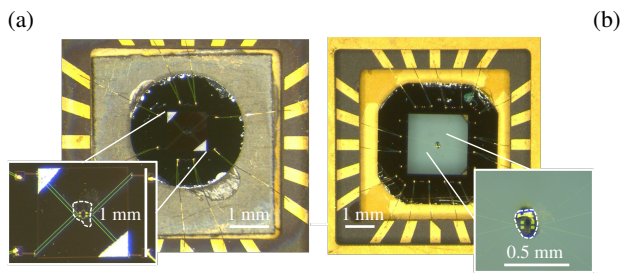

The thermal link between sample and bath is controlled by the membrane size, which can be tuned in order to control the experimental time constants.
This allows the system to be used for a range of sample sizes and heat capacities while retaining easily measurable time constants.
Two devices with different membrane sizes are shown in Fig.~\ref{fig:CBT pics}.
These were used for the experiments discussed in sections~\ref{sec:validation} and~\ref{sec:CeRh2As2} respectively.

The accessible temperature range is determined by the geometry of the junctions themselves.
Larger junction area gives a higher capacitance and thus lower charging energy, allowing access to lower temperatures.\footnote{But also lower resistance which can require using the low resistance theory and increase self-heating issues.}
Here we present results from \qty{0.04}{\micro\meter\squared} junctions used for low temperature measurements down to \qty{200}{\milli\kelvin}.
We have also produced \qty{0.004}{\micro\meter\squared} devices for measurements above \qty{1}{\kelvin}  and \qty{0.12}{\micro\meter\squared} ones for ultra low temperature measurements.
Smaller overlaps are realized using chevron shaped islands,\cite{Meschke2016} while for larger junctions we use a rectangular design\cite{Kauppinen1996} (see Fig.~\ref{fig:ShadowMask_withSEM2}).
Based on preliminary studies an array of $3$ rows of $N=12$ junctions has been chosen to give a reasonable combination of dip width and overall device resistance.

\subsection{Calorimeter Fabrication}
To create the calorimeter, the thin film structures are fabricated on commercially available silicon chips with silicon nitride membrane windows.
Devices used for measurements in section \ref{sec:validation} were based on PELCO 21522 silicon chips with 1~\si{\mm} $\times$ 1~\si{\mm} $\times$ 200~\si{\nm} \ce{SiN_x} membranes (see Fig.~\ref{fig:CBT pics}a).
Measurements in \ref{sec:CeRh2As2} used custom undoped silicon chips with predeposited gold contact pads and a 2~\si{\mm} $\times$ 2~\si{\mm} $\times$ 200~\si{\nm} silicon nitride membrane from NORCADA (see Fig.~\ref{fig:CBT pics}b).

\begin{figure}[t]
    \centering
    \begin{overpic}[width=0.9\linewidth]{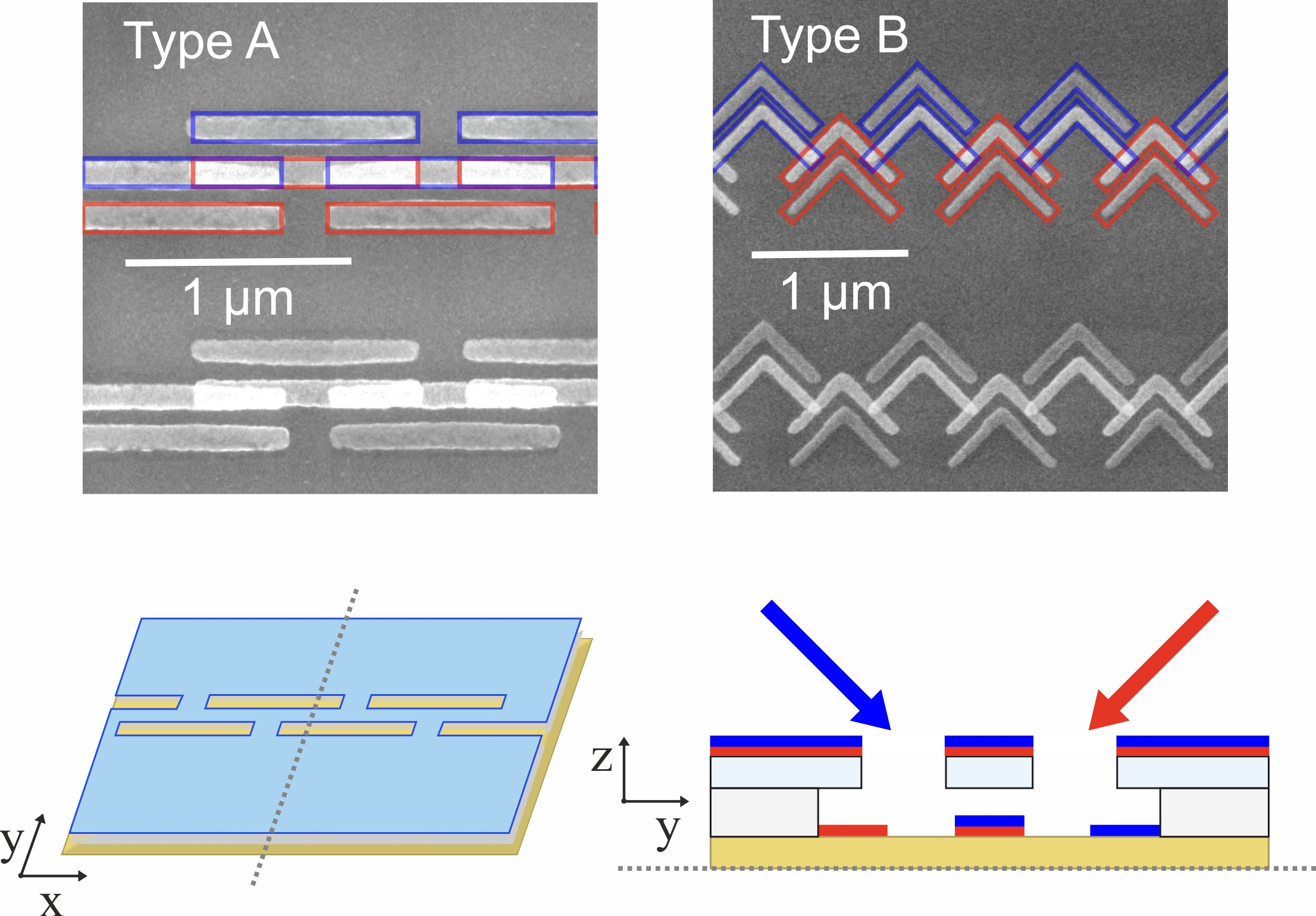}
    \put (0, 68) {(a)}
    \put (47, 68) {(b)}
    \put (0, 25) {(c)}
    \end{overpic}
    \caption{\textbf{a}: The CBT junction type A, realized from the mask shown in \textbf{c}. The red outlines indicate the first evaporation and the blue outlines indicate the second. \textbf{b}: The CBT junction type B, used for the high-temperature regime. \textbf{c}: An overview of the shadow mask evaporation technique. Left: The mask used to create the CBT junction array. Right: A cross section showing the overhanging top layer. The blue and red arrows indicate the two separate evaporation processes that occur at different angles, creating the overlapping structures in the center of the figure from which the tunnel junctions are formed. }
    \label{fig:ShadowMask_withSEM2}
\end{figure}

The key nanofabrication step is the realization of the CBT devices themselves.
For this we employ an electron beam lithography based shadow mask process.\cite{Dolan1977}
A schematic of the shadow mask is shown in Fig.~\ref{fig:ShadowMask_withSEM2}c.
The membranes are spin coated with a dual layer Copolymer PMMA/MA 33~\si{\percent} and PMMA 950K 2.5~\si{\percent} resist.
The Copolymer PMMA/MA \qty{33}{\percent} has a much higher sensitivity, allowing creation of the undercut necessary for the shadow evaporation technique.
The spin coating is carried out at \qty{6000}{rpm} for \qty{35}{\second} with subsequent baking for \qty{4}{\minute} on a hotplate at \qty{160}{\degreeCelsius} for each layer.

A JEOL JBX 6300FS system is used for the electron beam lithography steps.
Parameters were \qty{100}{\kilo\volt} acceleration voltage, \qty{100}{\pico\ampere} beam current, dose of \qty{1000}{\micro\coulomb\per\square\cm} for the main structures and \qty{200}{\micro\coulomb\per\square\cm} for the undercuts.
After development with a MIBK/Isopropanol 1:3 solution for \qty{90}{\second} this results in the mask cross section shown schematically in Fig.~\ref{fig:ShadowMask_withSEM2}c.

The aluminum evaporation and oxidation is carried out in a custom built UHV system.
A nominal initial 25~\si{\nano\meter} layer of aluminum is deposited under UHV at \qty{-30}{\degreeCelsius} and at an angle of approximately \qty{14}{\degree} to the sample normal, achieved by tilting the sample relative to the incoming \ce{Al} flux.
As a result through the resist windows we obtain the red outlined \ce{Al} structure on the substrate shown in Fig.~\ref{fig:ShadowMask_withSEM2}.
With the resist left in-place the chip is then transferred into the connected oxidation chamber and exposed to 0.15~\si{\milli\bar} oxygen pressure for approximately 20 minutes, with the oxidation monitored with a quartz oscillator which was covered with a fresh layer of \ce{Al} in parallel to the sample.
The sample is returned to the deposition chamber, cooled back to \qty{-30}{\degreeCelsius} and tilted to \qty{-14}{\degree}.\footnote{As the system only allows tilting in a single direction this is achieved by rotating the sample \qty{180}{\degree} about the $z$ axis then tilting by \qty{14}{\degree}.}
A second evaporation of nominally \qty{30}{\nano\meter} of \ce{Al} leads to the blue-outlined \ce{Al} structure shown in Fig.~\ref{fig:ShadowMask_withSEM2}. The red and blue-outlined structures partly overlap, creating the thin oxide in between which act as the tunnel junctions of the device. Please note, the cooling during Al evaporation allows for a smooth Al film, the monitored growth of the oxide layer to excellent reproducibility of the oxide thickness.
The mask and excess aluminum are removed using a 75~\si{\degreeCelsius} bath of NEP for three hours.
Following fabrication the chips are mounted on the chip carrier with epoxy.
After bonding between carrier and chip  all chip carrier pads are bonded together to prevent the generation of any potentially damaging voltages / electrostatic discharging across the CBTs during transport.

SEM images after the removal of the resist are shown in Fig.~\ref{fig:ShadowMask_withSEM2}.
This shows both the low temperature rectangular islands used for the results presented here (Fig.~\ref{fig:ShadowMask_withSEM2}a) as well as high-temperature chevron islands that enabled the creation of small junctions for measurements above \qty{1}{\kelvin} (Fig.~\ref{fig:ShadowMask_withSEM2}b).
Samples can be mounted to the rear of the membrane through an access hole in the chip carriers.

\subsection{Cryogenic Environment and Measurement Electronics}
\begin{figure}[tb]
    \centering
    \includegraphics[width=1\linewidth]{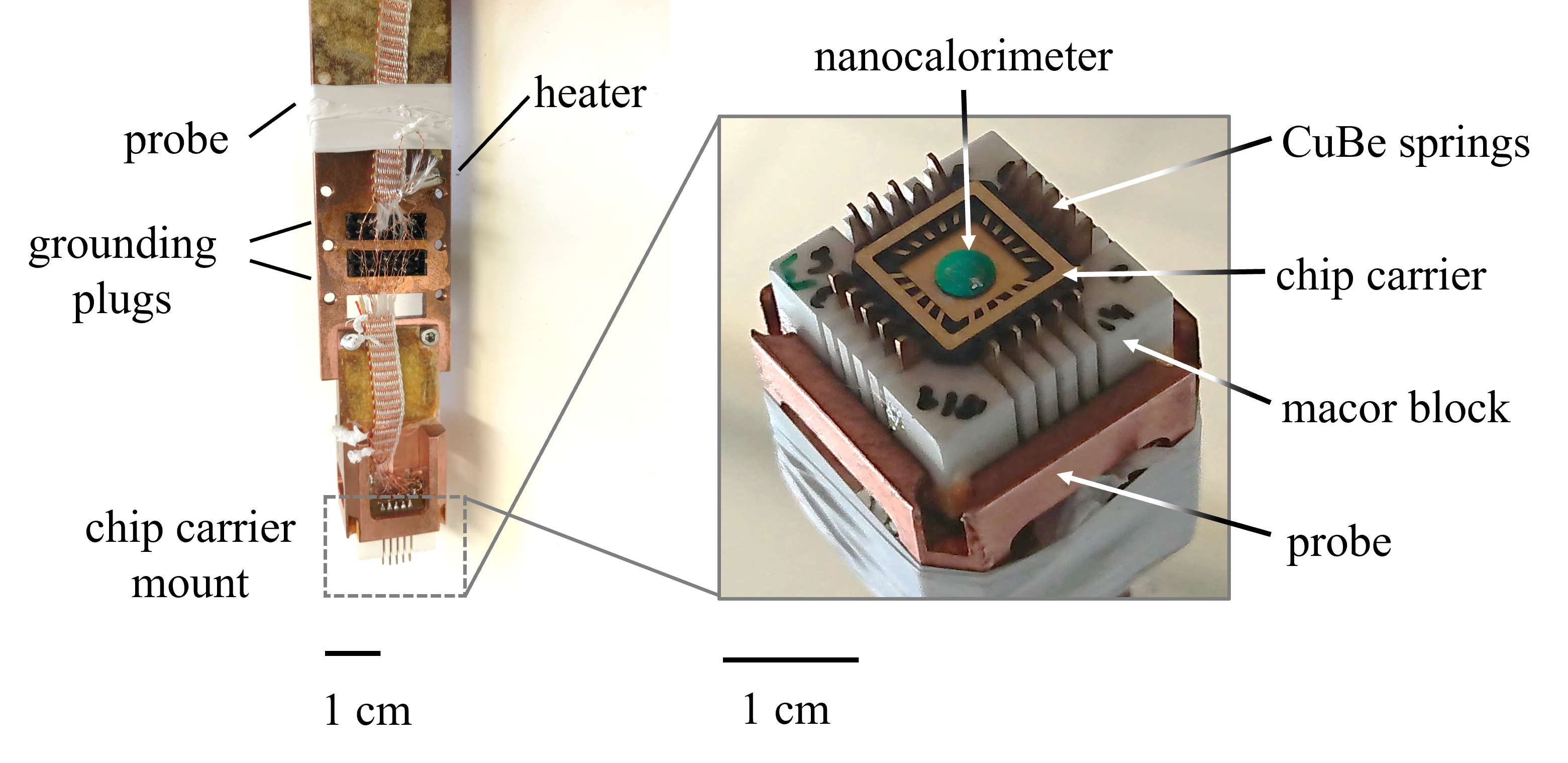}
    \caption{The CBT-speciﬁc measurement probe which attaches to the dilution refrigerator.
    The CBT chip carrier is mounted into the holder at the bottom of the probe as shown in the magnified view.
    The \ce{BeCu} springs and macor block ensure good thermal contact between chip carrier and probe.
    When in place, the grounding plugs short-circuit all of the measurement lines together to ensure safe handling of the CBTs until the probe is connected to the external grounding switch.
    }
    \label{fig:probe}
\end{figure}

The calorimeter interfaces with the cryostat via a custom probe shown in Fig.~\ref{fig:probe}.
The chip carrier is held in place on a Macor block using \ce{BeCu} springs in order to provide good electric and thermal contact.
The oxygen free high conductivity copper based probe is equipped with a \ce{RuO2} heater and a grounding plug to protect the CBTs against accidental discharges during installation. 

\begin{figure}
    \centering
    \includegraphics[width=1\linewidth]{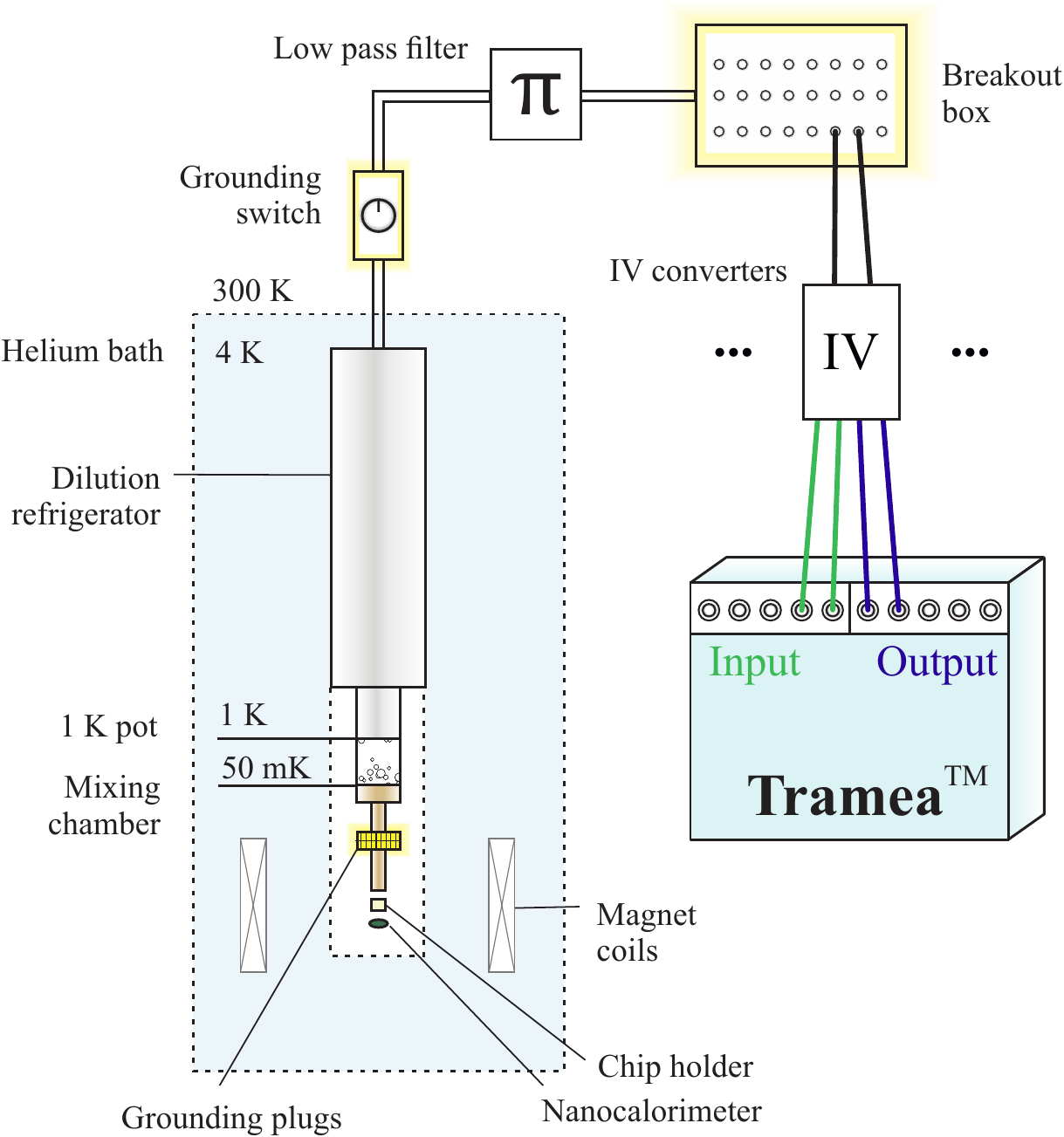}
    \caption{The experimental set-up, with the key components for CBT protection highlighted in yellow.
    The nanocalorimeter is mounted to the chip holder on the measurement probe.
    The \textit{grounding plugs} short-circuit all measurement lines together, and are removed once the probe is connected to the dilution refrigerator.
    The \textit{grounding switch} is connected to the dilution refrigerator measurement lines and it forms the principal protection for the sensitive CBT devices.
    The \textit{filter box} has a \(\pi\)-filter per measurement line.
    The \textit{breakout box} splits the single 24-line measurement cable into individual BNC connections and CBTs can be grounded individually. 
    The CBTs are measured using \textit{IV converters}, one per CBT.
    All measurements are controlled by the \textit{Tramea}, a specialist measurement unit with multiple channels and lock-in amplifiers.}
    \label{fig:design overview}
\end{figure}

The probe is mounted onto the mixing chamber of a dilution refrigerator with a base temperature of \qty{50}{\milli\kelvin}. Thermal contact is ensured by six annealed silver wires of \qty{1}{\milli\meter} diameter between the probe head and the mixing chamber. The cryostat is equipped with a \qty{1}{\tesla} / \qty{1}{\tesla} / \qty{9}{\tesla} vector magnet. (see Fig.~\ref{fig:design overview} for details)

In order to protect the CBTs in between measurement runs a grounding switch (Elma 04-4624) is mounted directly on the top of the cryostat.
This is followed by a \qty{1}{\kilo\hertz} low pass  \(\pi\)-filter to prevent parasitic heating by high frequency electrical noise.
Individual measurement lines can be grounded and interfaced at a breakout box. 

\begin{figure}[htb]
    \centering
    \begin{overpic}[width=0.8\linewidth]{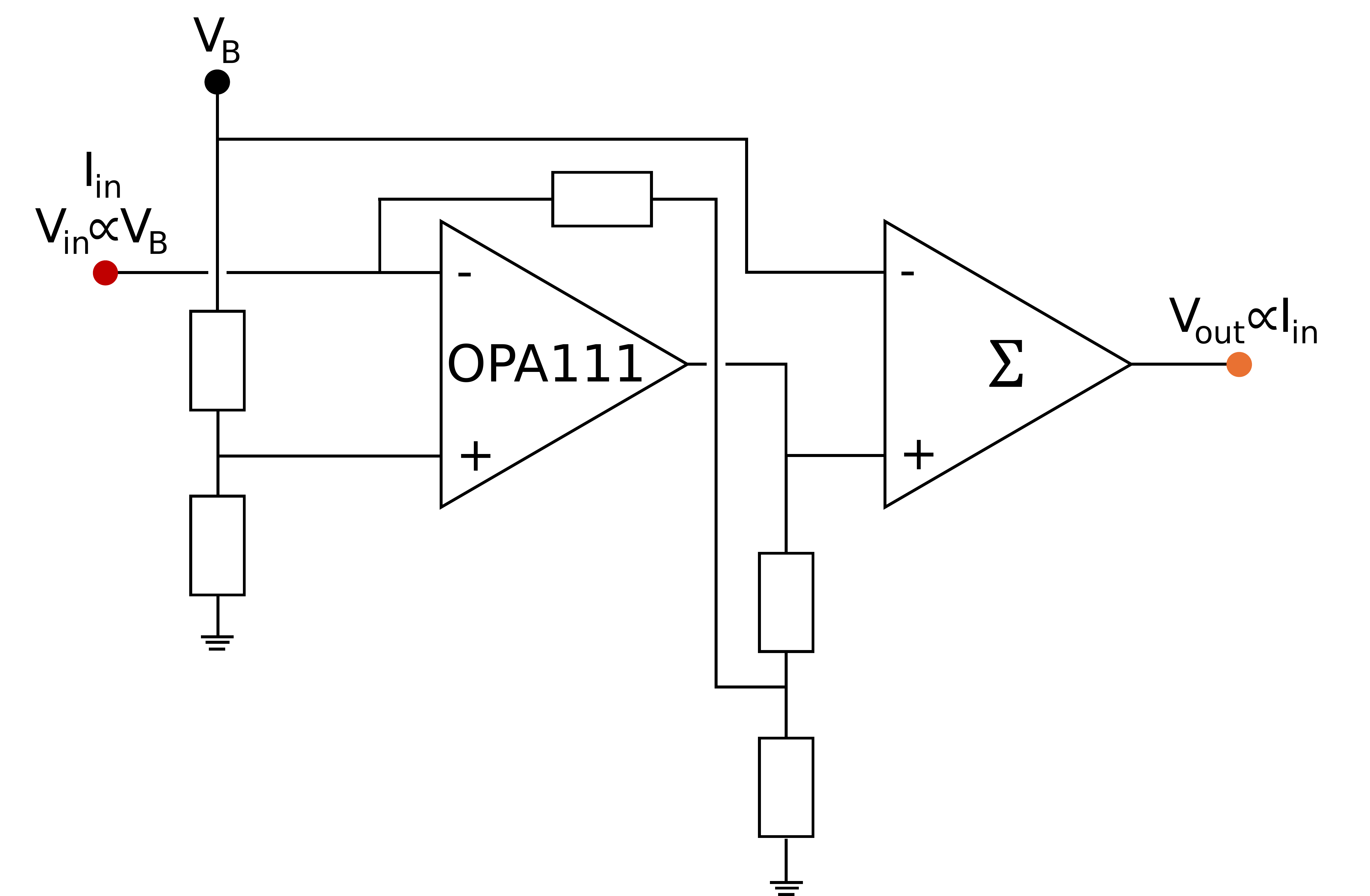}
        \put (2, 65) {(a)}
    \end{overpic}
    \begin{overpic}[width=0.8\linewidth]{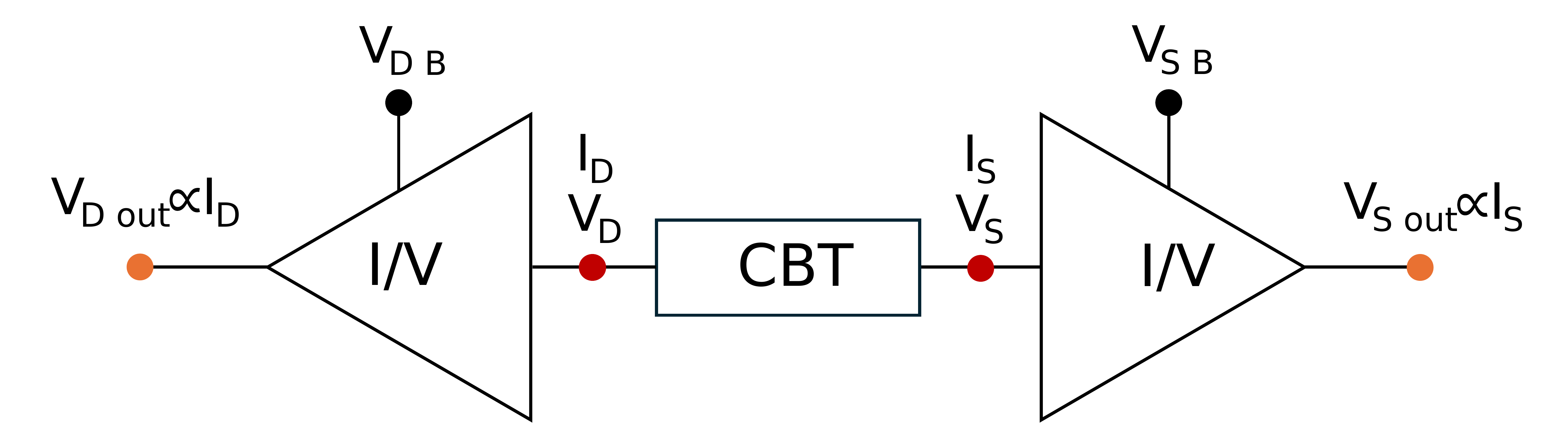}
        \put (2, 30) {(b)}
    \end{overpic}
    \caption{A schematic of the working principle of the home-built IV converters used to measure the CBT signal. \textbf{a}: the internal layout of the IV. \textbf{b}: A schematic layout of the measurement. The source (S) and drain (D) are connected to separate inputs and outputs.}
    \label{fig:IV converter schematics}
\end{figure}

In order to achieve optimal low noise measurements the CBTs are operated via home-built IV converters that combine bias voltage and IV-converter capabilities within one compact and properly shielded environment. 
A schematic of the electronic circuit is shown in Fig.~\ref{fig:IV converter schematics}. 
A $\pm$\qty{10}{\volt} input signal is converted to a bias voltage of $\pm$\qty{39}{\milli\volt}.
he IV converter itself is based on OPA111BM op-amps and provides a \qty{2.5E7}{\volt\per\ampere} amplification,\footnote{These are based on IV converters developed and used for many years for the characterisation of single-electron devices (for example quantum dots\cite{Weis1993}) in the group of J{\"u}rgen Weis at the Max Planck Institute, Stuttgart} used already for  many years.\cite{Weis1993}
For large bias voltages a commercial variable gain IV converter (Femto DLPCA-200) is used. However it was found that it causes significant parasitic heating  below \qty{1}{\kelvin}. 

The IV converters are controlled and managed by a SPECS Tramea\textsuperscript{TM} multi-channel measurement unit.
Lock-in modules as well as built-in bias sweepers with a \qty{400}{\kilo\hertz} bandwidth are used to directly obtain the differential conductance of the CBTs.
For ultra-low temperature measurements additional filtering of the Tramea outputs using CLPFL-0010-BNC \qty{10}{\mega\hertz} low pass filters was implemented.

\subsection{\label{sec:characterisation}Device characterisation}

\begin{figure}
    \centering
    \begin{overpic}[width=1\linewidth]{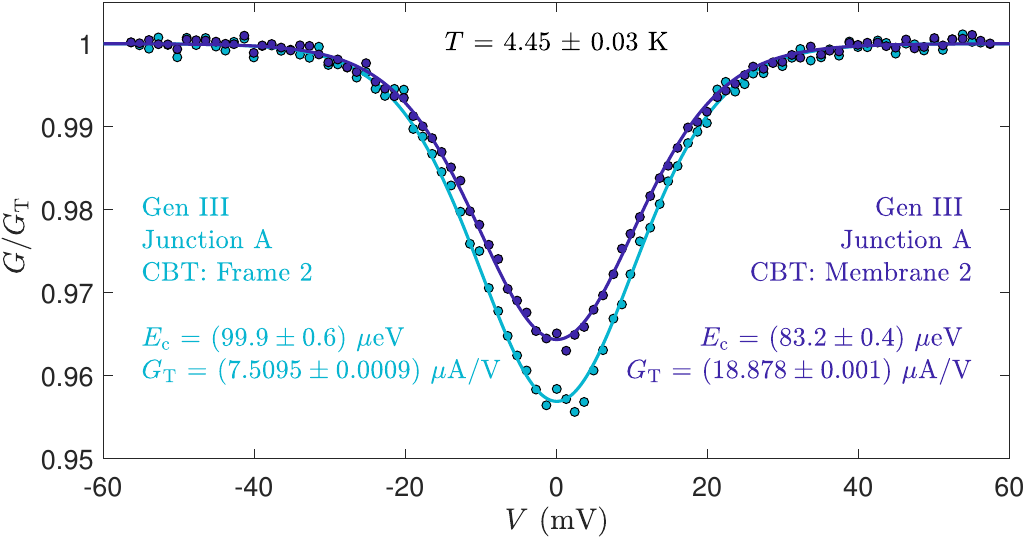}
        \put (0, 53) {(a)}
    \end{overpic}
    
    \begin{overpic}[width=1\linewidth]{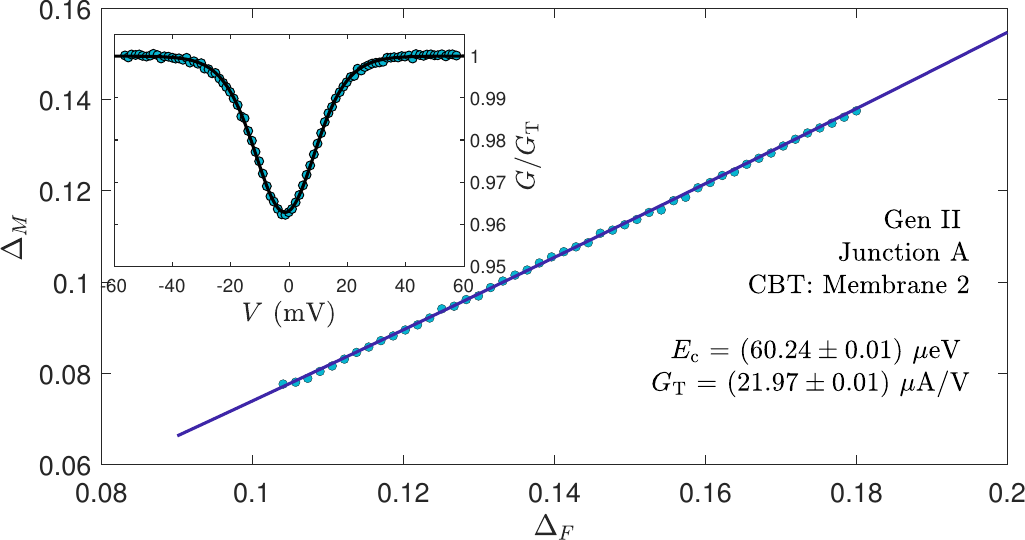}
        \put (0, 53) {(b)}
    \end{overpic}
    \caption{Examples of calibration dips. Dots show experimental data while lines are fits to the low resistance theory. \textbf{a}: Dips used for calibration of both Frame and Membrane CBTs of the third generation chip used for the \ce{CeRh2As2} study. \textbf{b}: The \(\Delta_M-\Delta_F\) plot to calibrate the membrane CBT of the second generation chip used for validation. Inset: Dip used to calibrate the frame CBT.}
    \label{fig:calibration dips}
\end{figure}

In Fig.~\ref{fig:calibration dips}a we present typical primary mode calibration data for a frame (cyan) and membrane (purple) CBT. During the measurement the inner vacuum chamber was flooded with helium exchange gas ensuring good thermal contact between the thermometers. 
The resulting temperature of \((4.45\pm0.03)\)~\si{\kelvin} is consistent with the helium bath conditions under which the measurement was taken.
Both data and fits are shown, with the quality of the fit confirming a constant CBT temperature throughout.

In Fig.~\ref{fig:calibration dips}b  we show an example of a \(\Delta_M-\Delta_F\) calibration on a second device used in cases where self-heating on the membrane does not allow for direct calibration due to the finite bias voltages required. In the insert we show the calibration for the well thermalized  frame thermometer. The main figure shows the plot of the conductance dip of the membrane thermometer \textit{vs} the frame thermometer together with the calibration fit discussed above (for more details see Appendix~\ref{ap:LR cal}). Calibration values for both devices used in the work presented here are shown in table~\ref{tbl:calibrations}.

\begin{table}[b]
    \centering
    \begin{tabular}{l|cccc}
        \hline
        \hline
        CBT & $G_T$& $R_J$& $E_C$& $E_C$\\
         & (\si{\micro\ampere\per\volt}) & (\si{\kilo\ohm}) & (\si{\micro\electronvolt})& (\si{\milli\kelvin})\\
        \hline
        Gen II Frame & $28.545\pm0.002$ & 8.8 & $85.2\pm0.4$&$988$\\
        Gen II Membrane & $21.97\pm0.01$ & 11 & $60.24\pm0.01$&$699$\\
        \hline
        Gen III Frame & $7.5095\pm0.0009$ & 33 & $99.9\pm0.6$& $1159$\\
        Gen III Membrane & $18.878\pm0.001$  & 13.3 & $83.2\pm0.4$&$965$ \\
        \hline
    \end{tabular}
    \caption{Calibration values for the calorimeters used in the work presented here. Both used the low temperature theory. The approximate junction resistance and charging energy temperature are also given for reference. }
    \label{tbl:calibrations}
\end{table}

\section{Measurement Modes and \label{sec:simulation}use of Numeric Simulations}

There are two standard experiments used to measure heat capacity, DC relaxation\cite{Bachmann1972} and steady state AC.\cite{Sullivan1968}

For relaxation measurements a square wave heating is applied with sufficiently low frequency to allow the sample to reach equilibrium between heater power changes.
The equilibration occurs exponentially with time constant $\tau=C/K$ with  specific heat $C$ and the thermal link to the bath $K$, which in turn can be calculated from the equilibrium temperature change induced by the varying heater power:
\begin{equation}
\label{eqn:DC responce}
    T = T_0 + \Delta T e^{-t/\tau}\;.
\end{equation}

An AC excitation instead induces a temperature oscillation.
At low frequencies this is essentially decoupled from the sample's heat capacity and to first order depends only on the ratio of the applied power and thermal link.
Above the cutoff frequency ($1/\tau$) the heat capacity starts to dampen the oscillations.
In this regime the product of frequency and oscillation amplitude is constant giving the quasi-adiabatic plateau, with the amplitude given by the ratio of the excitation power to the sample's heat capacity.

\begin{figure}[tb]
    \centering
    \begin{overpic}[width=\linewidth]{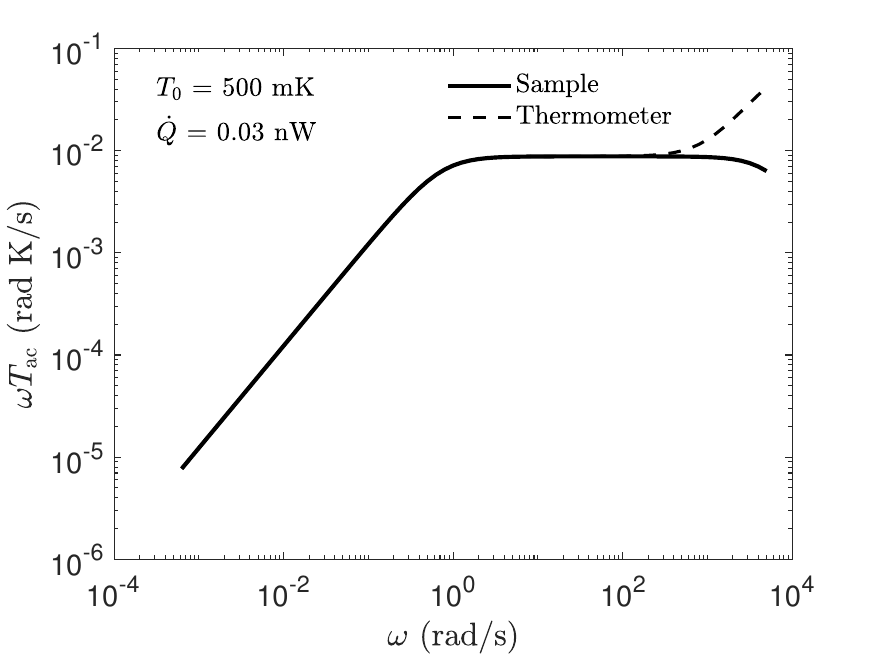}
        \put (33, 6) {\includegraphics[width=0.28\textwidth]{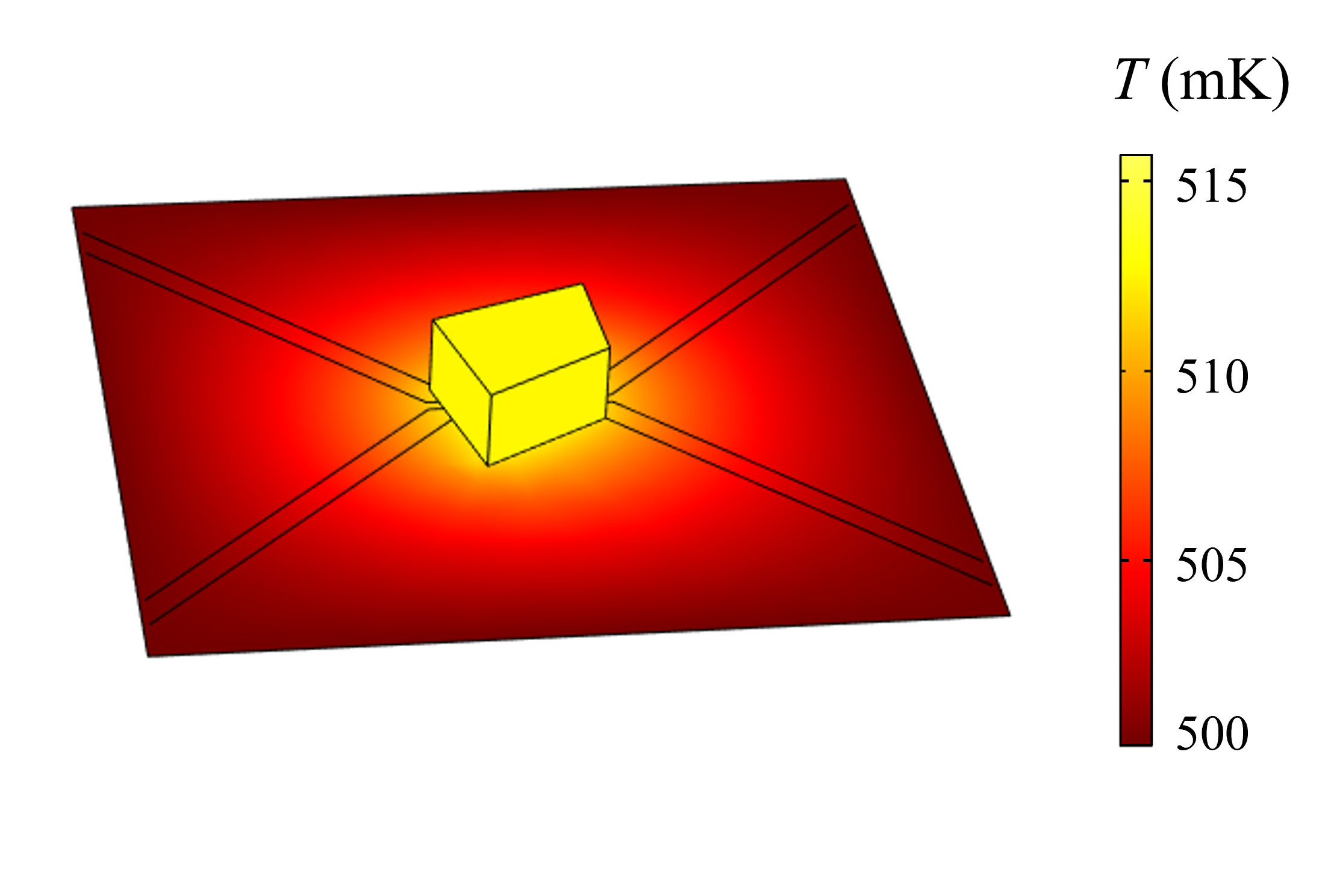}}
    \end{overpic}
    \caption{Simulation of the response of sample and thermometer to AC heating. Due to the geometry of the setup the thermometer decouples from the sample at high frequencies, instead following the heater. Inset: An example of the thermal profile across the membrane during a simulated heat pulse.}
    \label{fig:sr327-freqsweep-prediction}
\end{figure}

In Fig.~\ref{fig:sr327-freqsweep-prediction} we show the characteristic frequency response for our measurement geometry in terms of the amplitude of the temperature oscillations $\tilde{T}$ as a function of $\omega = 2 \pi f$ where $f$ is the frequency of the applied ac heating with amplitude $P_0$. Mathematically it is given by 

\begin{equation}
    \label{eqn:AC responce}
    \tilde{T} = \frac{P_0}{\omega C}\left(1 + \frac{1}{\left(\omega\tau\right)^2} + \zeta\left(\tau_i, K_s, K\right)\right)^{-1/2}
\end{equation}

with $\zeta$ accounting for correction factors from the internal time constant $\tau_i$ and thermal conductance $K_s$ of the sample as well as the thermal link to the bath $K$. In a well designed setup  these are typically small and geometry dependent.
For nanocalorimetry the two-dimensional geometry of the calorimeter as well as sample shape may not conform to the conditions of simple analytic models, requiring complex additional analytical terms to accurately describe the calorimeter behavior.\cite{Tagliati2011}
Here state-of-the-art simulations can be used to quantify the necessary corrections.
For this purpose, and to assist the design of the calorimeter in general, we have used finite element method simulations using COMSOL Multiphysics,\textsuperscript{\textregistered{}}\cite{COMSOL} performing AC and DC heat capacity measurements \textit{in silico}. 

By comparing the 'input' heat capacity defined in the software and the 'output' heat capacity determined from the simulated measurement, using identical data processing to the physical experiment, a numerical correction factor can be determined and applied to experimental data.
This also allowed prediction of thermal conductance geometry factor as the highly 2D geometry of the membrane makes analytically determining this challenging.
We calculated a geometry factor of \qty{1.3E6}{\per\meter}, with a literature conductivity of \qty{0.0033}{\watt\per\meter\per\kelvin} at \qty{500}{\milli\kelvin}\cite{Zink2004} giving a thermal link of 2.5~\si{\nano\watt\per\kelvin}.
The conductivity of \ce{SiN_x} membranes does however vary significantly depending on their exact composition and tension\cite{Zink2004, Holmes1998} so this conductance is not necessary representative of our devices.
Fig.~\ref{fig:sr327-freqsweep-prediction} shows an example of the simulation being used to predict the frequency dependence for AC measurements, where the atypical geometry of the device means the thermometer decouples from the bulk of the sample before the sample decouples from the heater.

For a well placed sample covering both the heater and thermometer, the difference between input and output heat capacities we found to be small for our devices, at most 1~\si{\percent}.
Corrections can also be calculated for a non-ideal sample placement, thereby accounting for sub-optimal measurement conditions numerically rather attempting to move the sample, risking destroying the fragile membrane, see Appendix~\ref{ap:offcenter-correction}.

\section{\label{sec:validation}Calorimeter Validation}

\begin{figure}
    \centering
    \includegraphics[width=1\linewidth]{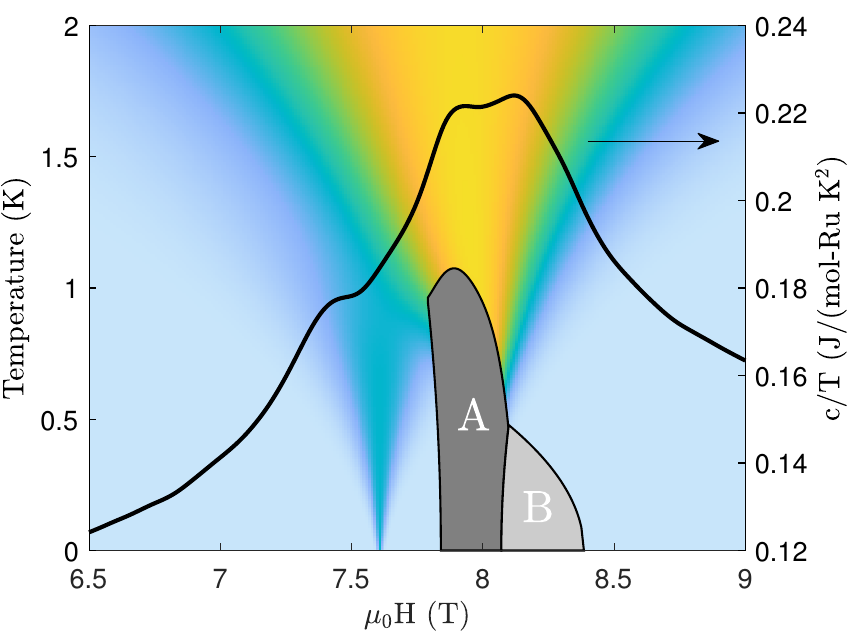}
    \caption{The low temperature \(T - H\) phase diagram for \ce{Sr3Ru2O7}, for \(H\parallel c\).
    Phase A is bounded by two first order transition lines at \(H_1\), \(H_2\), with a second order transition line forming the “roof” of the phase. The color shows a schematic representation of the area of influence of the quantum critical end points, with the specific heat at \qty{400}{\milli\kelvin} shown on top.\cite{Rost2010}}
    \label{fig:327 phases}
\end{figure}

To validate the calorimeter performance we measured the heat capacity of a microcrystal of the itinerant metamagnet \ce{Sr3Ru2O7}.\cite{Bruin2013}
Below \qty{1}{K} and at high magnetic fields, the relevant regime for our calorimeter, this material has a well mapped complex phase diagram, as shown in Fig.~\ref{fig:327 phases},\cite{Sun2018, GRIGERA2001, GRIGERA2004, Borzi2007, Rost2009a, Rost2009b} together with an example of the evolution of the specific heat as a function of magnetic field at \qty{400}{\milli\kelvin}.
Up to \qty{7}{\tesla} \ce{Sr3Ru2O7} shows Fermi liquid behavior in the relevant temperature regime with $C/T$ varying from 0.1 to \qty{0.15}{\joule\;mol-Ru^{-1} K^{-2}}. At \qty{7.5}{\tesla} a thermodynamic cross over occurs followed by two first order transitions marking entry and then exit of phase A. At low temperatures an additional phase B sits between A and the high-field Fermi liquid regime, but it has an extremely weak signature in specific heat and is difficult to resolve.\cite{Borzi2007, Bruin2013, Lester2015} The indicated "roof" of phase A is a second order phase transition into a high-temperature regime at approximately \qty{1}{K} that does not show conventional Fermi liquid behavior.\cite{GRIGERA2001}

The well established properties and key features of these phase transitions as well as the presence of Fermi liquid regimes make \ce{Sr3Ru2O7} an ideal characterisation material. For our measurements we used an off-cut from a larger well characterized crystal (C698K\cite{Mercure2008, Rost2009b, Mercure2010}). Determination of the sample mass is one of the most significant challenges of performing nanocalorimetry with absolute accuracy. Here we use magnetic susceptibility measurements\cite{QDMPMS2019, QDPPMS2019} as a function of temperature to estimate the sample amount to be $(6.69\pm0.68)\times10^{-8}$~\si{\mole\of{Ru}} corresponding to $(19.3\pm2.0)$~\si{\ug}. This sample mass is more than a thousand times smaller than the 24.1 mg crystal measured to produce the existing specific heat data set for \ce{Sr3Ru2O7}.\cite{Rost2009a, Rost2009b, Rost2011} At \qty{550}{\milli\kelvin} and \qty{5}{\tesla} the heat capacity of the sample is estimated\cite{Rost2009b} to be \qty{4}{\nano\joule\per\kelvin} with the typical thermal conductance of the membrane estimated to be \qty{5}{\nano\watt\per\kelvin}.

The calorimeter used has a \qty{200}{\nm} thick, \qty{1}{\mm} wide square membrane. The estimated thermal conductance (see section \ref{sec:simulation}) results in a characteristic time constant of $\tau=C/K=0.8$~\si{\second} comparable to previous conventional experiments on \ce{Sr3Ru2O7}.\cite{Rost2010} The membrane / frame thermometer characteristics are discussed in section \ref{sec:characterisation} with a charging energy of \((60.24 \pm 0.01)\)~\si{\micro\electronvolt} / \((85.2 \pm 0.4\))~\si{\micro\electronvolt}, and infinite conductance \(G_T\) of \((21.97 \pm 0.01)\)~\si{\micro\ampere\per\volt} / \((28.545 \pm 0.01)\)~\si{\micro\ampere\per\volt} respectively. The base temperature of this device is 300~\si{\milli\kelvin}.

\begin{figure}[htb]
    \centering
    \includegraphics[width=\linewidth]{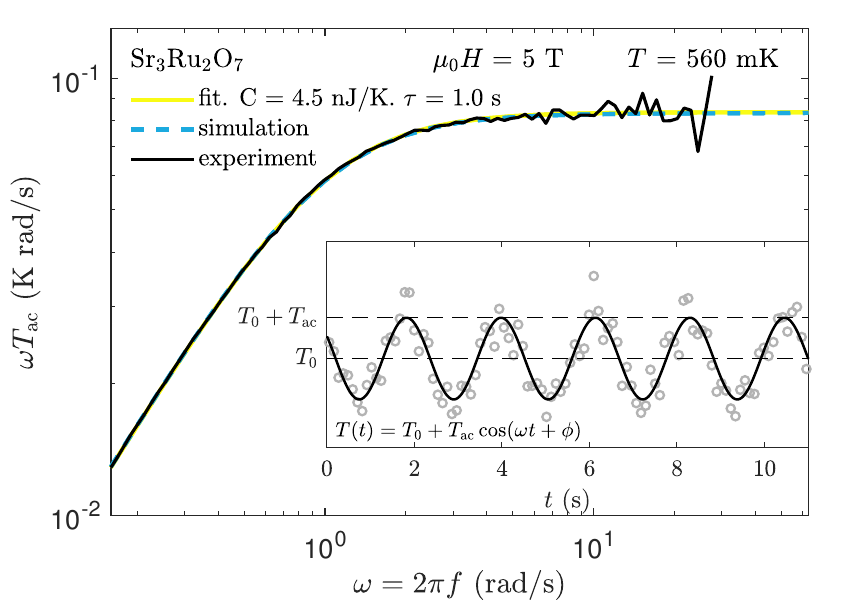}
    \caption{A comparison of the measured frequency response of an AC method measurement of \ce{Sr3Ru2O7} with a numerical simulation. The dashed line shows a fit to equation~\eqref{eqn:AC responce} confirming the expected frequency dependence. Inset: An example time trace for an AC method measurement. The solid black line is the fitted curve used to extract the oscillation amplitude.}
    \label{fig:sr327-freqsweep}
\end{figure}

In Fig.~\ref{fig:sr327-freqsweep} we show in black the AC response of the sample thermometer as a function of $\omega=2\pi f$. The applied ac excitation of the heater is $P_0=0.3725$~\si{\nano\watt}. 
The dashed line shows a fit of equation~\eqref{eqn:AC responce} with $\zeta=0$ to the data resulting in a heat capacity $C=$\qty{4.5}{\nano\joule\per\kelvin} and time constant \qty{1}{\second}, giving $\kappa=$ \qty{5.85}{\milli\watt\per\meter\per\kelvin} derived from the thermal conductance $K=$ \qty{4.5}{\nano\watt\per\kelvin} with the geometry factor discussed in section \ref{sec:simulation}.
This thermal conductivity is in good agreement with previous reports on \ce{SiN_x} membranes in the range from $1$ to $10$ \si{\milli\watt\per\meter\per\kelvin} given the fabrication process dependent properties of such membranes.\cite{Zink2004, Holmes1998} The blue line shows the predicted response from COMSOL\textsuperscript{\textregistered{}} using the parameters from the fit.
The overall agreement between the three curves in Fig.~\ref{fig:sr327-freqsweep} is excellent. This confirms that the simulation model accurately describes the experimental setup and that equation~\eqref{eqn:AC responce} with $\zeta=0$ captures all relevant parameters in this frequency regime

\begin{figure}[htb]
    \centering
    \includegraphics[width=1\linewidth]{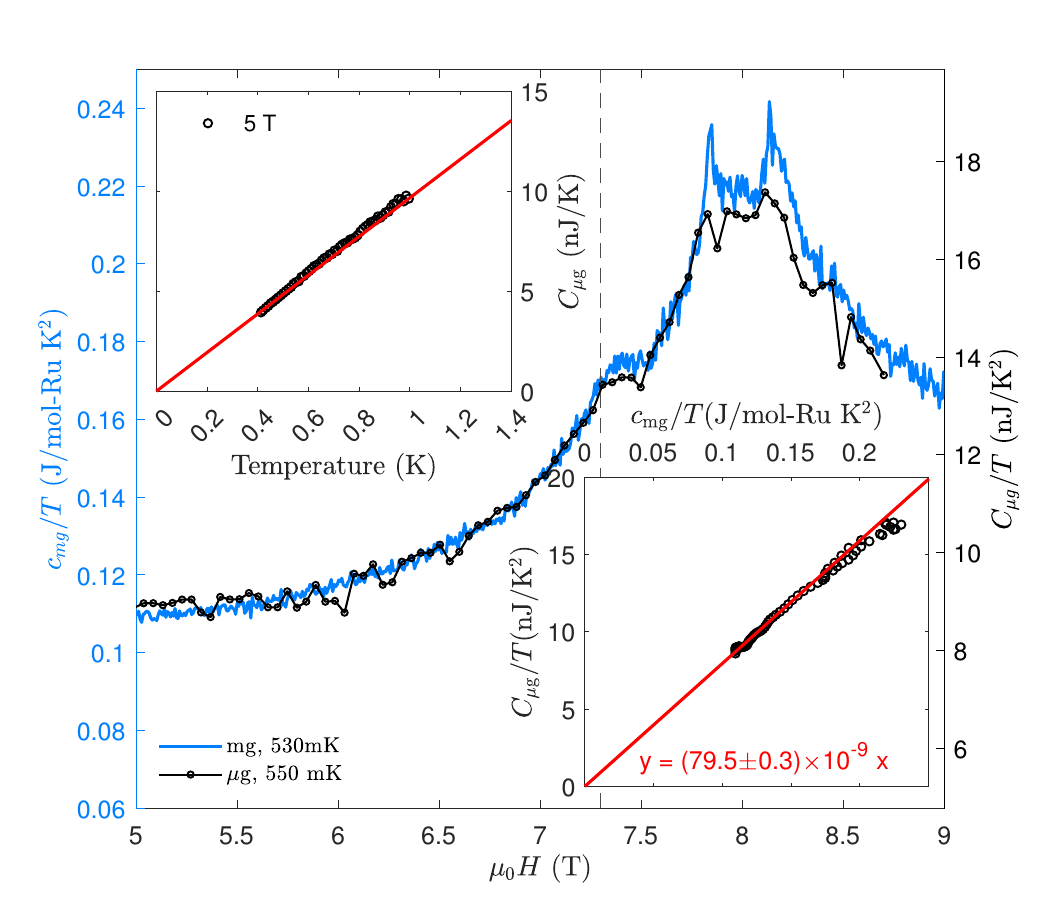}
    \caption{A comparison of our measured heat capacity (black with circular markers, right axis) with the literature specific heat data\cite{Rost2009b} (blue line, left axis, common zero) showing a high degree of agreement. In the Fermi liquid regime below 7.1 T heat capacity should scale linearly with temperature, which is the case as shown in the inset top left. In this regime specific heat divided by temperature of the measurement and literature sample are the same (despite the different measurement temperatures) allowing calibration of the sample mass as shown in the inset bottom right.}
    \label{fig:327 specific heat comparison}
\end{figure}

In Fig.~\ref{fig:327 specific heat comparison} we show the measured heat capacity divided by temperature $C_{\mu g}/T$ as a function of magnetic field at \qty{550}{\milli\kelvin} using a calorimetry measurement ($P=0.373$~\si{\nano\watt}, $\omega=3.1$~\si{\radian\per\second}) together with previous specific heat data divided by temperature $c_{mg}/T$ which at $T=$\qty{530}{mK} was taken at effectively the same temperature.\cite{Rost2009b, Rost2011} Note while the data are shown on different axes the two scales share a common zero.
The qualitative agreement between these is excellent, with the three transitions at $H_0, H_1, H_2$ visible in the microcrystal curve. For magnetic fields below 7.1~\si{\tesla} \ce{Sr3Ru2O7} behaves as a Fermi liquid resulting in \(C/T\) being temperature independent to first order.\cite{Rost2010, Rost2009a} In the upper left inset in Fig.~\ref{fig:327 specific heat comparison} we show $C_{\mu g}$ as a function of temperature at 5 T together with a linear guide-to-the eye highlighting the expected linearity of the heat capacity.
In the inset bottom right we plot heat capacity of the microcrystal against the previously established specific heat for the two field-dependent curves shown in the main figure. In the Fermi liquid regime up to 7.1 \si{\tesla} these are expected to be directly proportional to each other as confirmed by the linear fit shown in red. This linearity can be used to extract a sample amount, of \((79.5\pm0.3)\)~\si{\nano\mole\of{Ru}}. The difference of approximately \qty{20}{\percent} is within the systematic uncertainties of the susceptibility measurement on such a small sample and is comparable to those reported in other nanocalorimeter experiments.\cite{Tagliati2012}
Applying this mass correction scales our data onto the literature values, with a residual rms error between the two of just \qty{2.3}{\percent} of the literature heat capacity while the heat capacity varies by \qty{122}{\percent} within this regime.

In the data shown we are able to resolve heat capacity changes as small as \qty{0.1}{\nano\joule\per\kelvin}.  
For the previous measurement on the 20 mg sample\cite{Rost2009b} this was \qty{2}{\nano\joule\per\kelvin}. Given that both field-sweep experiments were carried out over approximately the same duration (between two helium fills) this indicates at least a factor 20 increase in sensitivity of this prototype. 

\section{\label{sec:CeRh2As2}The benefits of studying microcrystals}

In order to highlight the benefits available when studying microcrystals we measured the specific heat of the topical material \ce{CeRh2As2}.
This is a highly unconventional heavy fermion superconductor which was recently found by Khim \textit{et al.} to exhibit multiphase superconductivity as a function of magnetic field.\cite{Khim2021} Additionally an unconventional normal state phase transition has been found to exist above the superconducting domes that has been hypothesized to mark the formation of a quadrupole density wave (QDW),\cite{Hafner2022, Semeniuk2023} though recent measurements have also shown signs of AFM ordering.\cite{Khim2025}
In previous reports the transition to the superconducting phase is significantly lower in specific heat measurements than in transport, suggesting a range of transition temperatures within each sample.\cite{Khim2021, Semeniuk2023}

We have measured a microcrystal of a larger crystal of the same generation studied by Semeniuk \textit{et al.} (growth 87700).\cite{Semeniuk2023}
Based on magnetic susceptibility the sample amount is \((5.0\pm0.2)\times10^{-8}\)~\si{\mole}.
The calorimeter used is based on a \qty{200}{\nm} thick, \qty{2}{\mm} wide square membrane.
The membrane and frame thermometer characteristics are discussed in section \ref{sec:characterisation} with charging energies of \((83.2 \pm 0.4)\)~\si{\micro\electronvolt} and \((99.9 \pm 0.6)\)~\si{\micro\electronvolt}, and infinite conductances \(G_T\) of \((18.878 \pm 0.001\))~\si{\micro\ampere\per\volt} and \((7.5095 \pm 0.0009)\)~\si{\micro\ampere\per\volt} respectively.
The base temperature of this design was 200~\si{\milli\kelvin}.

\begin{figure}
    \centering
    \includegraphics[width=0.9\linewidth]{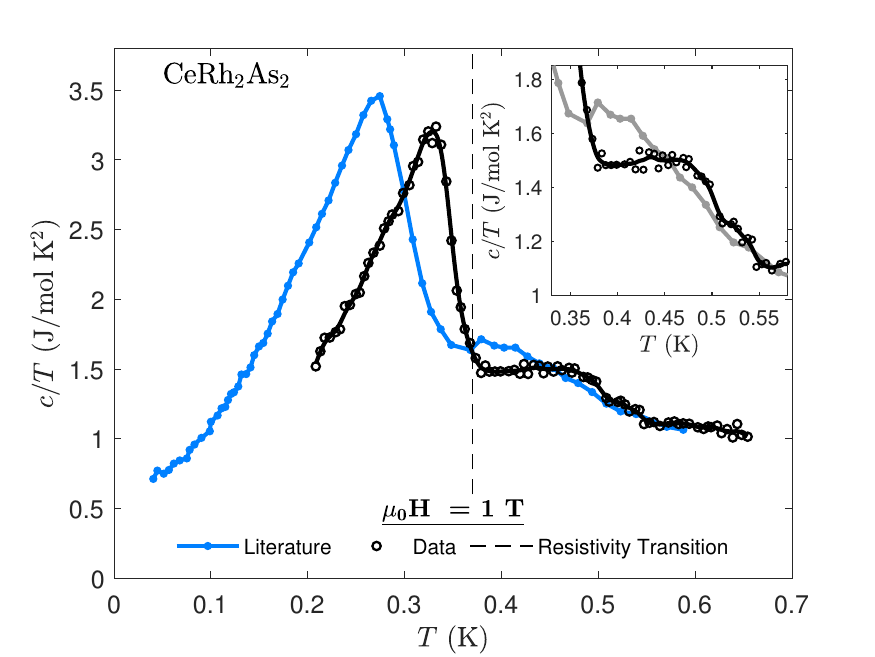}
    \caption{A comparison of the 1~\si{\tesla} measured \(C/T\) for the microcrystal, with the data published by Semeniuk et al.\cite{Semeniuk2023}
    Inset: A zoom in on the QDW transition for both curves.}
    \label{fig:CeRh2As2 CoT comparison}
\end{figure}

A plot of our data compared to that taken on a \si{\milli\meter} scale crystal by Semeniuk \textit{et al.} is shown in Fig.~\ref{fig:CeRh2As2 CoT comparison}.
Both sets were taken with a magnetic field of 1~\si{\tesla} applied along the crystallographic $c$-axis.
The transition widths of the microcrystal are significantly sharper than those of the larger crystal, with widths reduced by a factor 2-3 dependent on the criterion used and comparable to that observed in the next generation of samples.\cite{Khim2025} The onset of superconductivity in the microcrystal is furthermore consistent with the resistive transition temperature.\cite{Semeniuk2023}

The microcrystal measurement therefore enables the study of the phase diagram evolution with higher resolution than previously possible due to the sharper transition. In this material this is crucial to fully determine the nature of the interplay between QDW and superconductivity and will be explored in a future study.

\section{Conclusion}
We present a novel nanocalorimeter based on the inclusion of primary Coulomb blockade thermometry on a silicon nitride membrane.
The device is targeted towards \(\mu\)g scale samples, achieving a resolution below 0.1~\si{\nano\joule\per\kelvin} at 300~\si{\milli\kelvin}, allowing use with small pieces of unoptimized crystal and mesoscopic systems.
The magnetic field independent primary thermometry on the membrane removes systematic uncertainties in calibration and is well suited to diagnose parasitic heating, a major issue for highly isolated thermometers at \si{\milli\kelvin} temperatures.

The flexible processes used to construct the device allow for tuning of both thermal link and the arrangement of thermometers on the chip to suit a given set of experimental parameters.
Here we have presented a device optimized for \si{\milli\kelvin} heat capacity measurements but are also in the process of developing other arrangements for use with the Quantum Design PPMS\textsuperscript{\textregistered{}} as well as for thermal conductivity measurements.

As an example of the benefits of our new device we have shown measurements on a microcrystal of the unconventional superconductor \ce{CeRh2As2}.
Compared to the literature, studying a microcrystal we find significantly sharper transitions that are more consistent with transport measurements.\cite{Semeniuk2023}

\begin{acknowledgments}
The authors acknowledge the assistance in fabrication of measurement electronics by Maximilian K{\"u}hn and Lukas Freund at the Max Planck Institute, Stuttgart. We also wish to thank Robin S. Perry for providing the samples of \ce{Sr3Ru2O7} and Seunghyun Khim for the sample of \ce{CeRh2As2}.
We acknowledge funding from the Engineering and Physical Sciences Research Council [grants EP/P024564/1, EP/V049410/1 and EP/L015110/1] as well as the IMPRS-CPQM.
\end{acknowledgments}

\section*{AUTHOR DECLARATIONS}
\subsection*{Conflict of Interest}
The authors have no conflicts to disclose.

\subsection*{Data Availability Statement}
The data that support the findings of this study are openly available in the St Andrews Research Portal at \url{https://doi.org/10.17630/d282d2d0-6b23-460d-b951-50219735f147}.\cite{dataset}

\appendix

\section{Sample Mounting}
\label{ap:mounting}
Sample mounting is similar to that developed for small samples on delicate platforms such as torque cantilevers.
Sample were manipulated using a length of \qty{25}{\micro\meter} gold wire, held either in a pair of tweezers or using a micro-manipulator.
We first deposit a small spot of Apiezon N grease less than \qty{20}{\micro\meter} across in the center of the membrane to provide good thermal coupling.
We estimate the heat capacity of the grease to be on the order of \si{\pico\joule\per\kelvin},\cite{Schink1981} at least three orders of magnitude less than the sample and so completely negligible. 
Upon sample placement the grease covers the whole sample cross section and is therefore less than \qty{1}{\micro\meter} thick.
We have run simulations with the grease present\cite{Scheie2018} and have found it to provide a negligible thermal barrier.
The membranes are surprisingly resilient and were not damage during sample mounting.

\section{\label{ap:cal}Ensuring thermal equilibrium for $\Delta_M-\Delta_F$ calibration}

\begin{figure}[tbh]
    \includegraphics[width=0.9\linewidth]{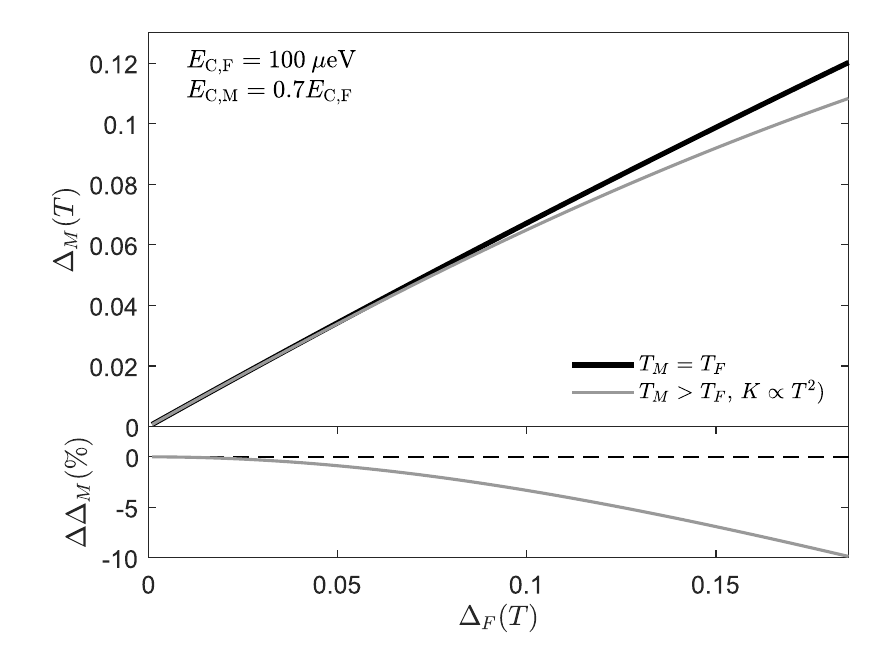}
    \caption{A simulated \(\Delta_M-\Delta_F\) plot showing the effect of the decoupling of membrane and frame due to parasitic heating and lowering thermal conductivities (\(T^2\) dependence of the thermal link).The bottom graph shows the difference of the two curves in the top graph.}
    \label{fig:example_delta_delta}
\end{figure}

Due to the well defined theoretical behavior of the CBTs as a function of temperature, any decoupling between frame and membrane CBTs is obvious as a deviation from the expected $\Delta_M-\Delta_F$ plot shape.
This is shown in Fig.~\ref{fig:example_delta_delta} where temperature independent parasitic heating leads to decoupling of the frame and membrane at low temperatures due to a decreasing thermal link. Unlike in the case of resistive thermometry this only limits the range usable for calibration but does not prevent measurement as the sample and membrane CBT remain well-coupled. It is thus possible to ensure calibration is only performed using data points where the frame and membrane CBT in thermal equilibrium at the same temperature.

\section{\label{ap:LR cal}Calibration With Low Resistance CBTs}

In the case of low junction resistance it can be challenging to achieve a high enough temperature so that \(u_R\) becomes much less than one and the first order equations become valid.
In this case the low resistance formulas must be used for calibration.
Fitting the dip remains fairly straightforward, though a more complex equation must be used it remains trivial for modern computers.
The low resistance dip is given by\cite{Farhangfar2001}

\begin{equation}
    G = G_T\left(1-\frac{e^2R}{\pi\hbar}\frac{N-1}{N}\frac{\partial{F\left(\nu, u_R\right)}}{\partial{\nu}}\right)\;,
\end{equation}

where \(\nu = \frac{eV}{Nk_BT}\) is the normalized voltage across the array , \(u_R = \frac{E_C}{4\pi^2 k_BT} \frac{R_K}{R_J}\) is the normalized charging energy, and \(F\left(\nu, u_R\right)\) is a function given by

\begin{multline}
    F\left(\nu, u_R\right) = \nu\left(\Re{\left[\Psi\left(1+u_R-\frac{i\nu}{2\pi}\right)-\Psi\left(1-\frac{i\nu}{2\pi}\right)\right]}\right.\\
    \left.-\Im{\left[2\pi u_R\Psi\left(1+u_R-\frac{i\nu}{2\pi}\right)\right]}\right) \;.
\end{multline}

\begin{figure}[h!]
    \centering
    \includegraphics[width=\linewidth]{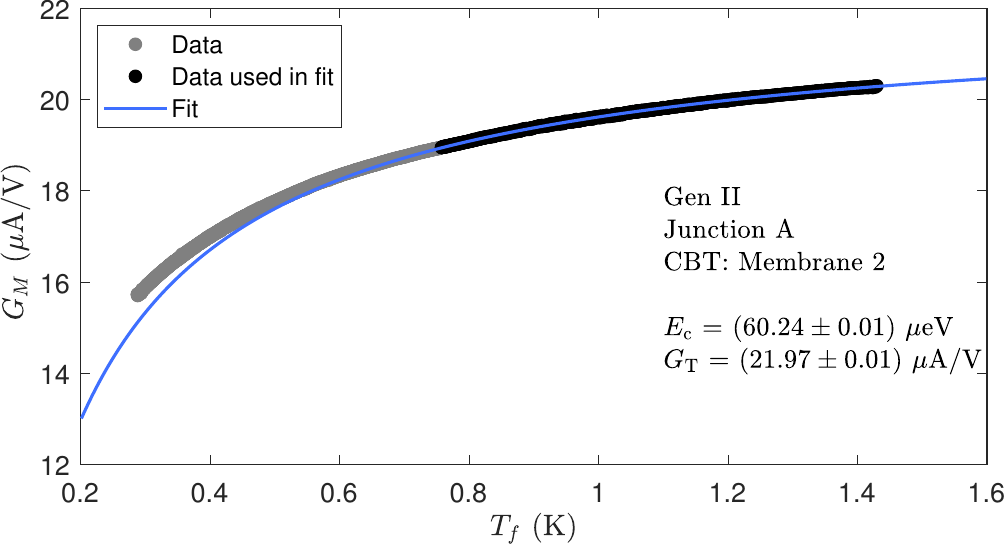}
    \caption{Calibration plot used to fit the second generation membrane CBT \(G_T\) and \(E_C\) using the known temperature of the frame CBT.
    Equilibrium between the two is maintained down to approximately 800~\si{\si{\milli\kelvin}} below which the membrane remains hotter, deviating from the high-temperature fit.
    }
    \label{ap:fig:G T fit}
\end{figure}

Due to the lack of an analytic inverse of equation~\eqref{eqn:CBT T LTLR DGOGT} performing the \(\Delta_M - \Delta_F\) fit directly is no longer straightforward.
Instead a calibrated frame thermometer can be used to fit the membrane thermometer conductance as a function of temperature, and a \(\Delta_M - \Delta_F\) plot can be constructed using this fit to check for thermal equilibrium.
As the frame \(\Delta\) and temperature are related by a single valued function this is in essence the same as the original \(\Delta_M - \Delta_F\) fit but computationally much easier.
Conveniently this also allows fitting of \(G_T\) in case fabrication defects on the membrane produce in-series resistances.
An example of a \(G_M - T_F\) fit is shown in Fig.~\ref{ap:fig:G T fit}, which was used to calibrate the membrane for the validation run.
Here low temperature data is excluded from the fit as the lack of exchange gas meant the membrane and frame were not always in equilibrium.
The \(\Delta_M - \Delta_F\) plot generated from this can be seen in the inset of Fig.~\ref{fig:calibration dips}a.
This remains straight over the temperatures used for the fit showing the frame and membrane to have been in equilibrium.

\section{\label{ap:offcenter-correction}FEM Corrections for off center samples}
Numerical analysis has proven especially valuable to correct for sample placement.
Ideally the sample would lie in the center of the membrane, fully covering both heater and thermometer CBTs and their thermalisation meanders.
However, due to the extreme fragility of the membrane moving the sample once placed poses a significant risk, so it is beneficial to be able to take measurements with the sample off center and, for example, covering only one of the CBTs (which would in this case be designated as thermometer and the uncovered CBT used as heater).
Measurements of the sample's placement can be taken using a microscope and recreated in the simulations, allowing a correction factor to be determined and applied to the experiment, greatly increasing accuracy.
Using this method geometry induced corrections of up to 25~\si{\percent} have been successfully accounted for.

Simple corrections are however limited to DC measurements as for AC measurements off center samples cause the sample and heater to decouple, making the correction factor sample heat capacity dependent.

\begin{figure}[h]
    \centering
    \includegraphics[width=1\linewidth]{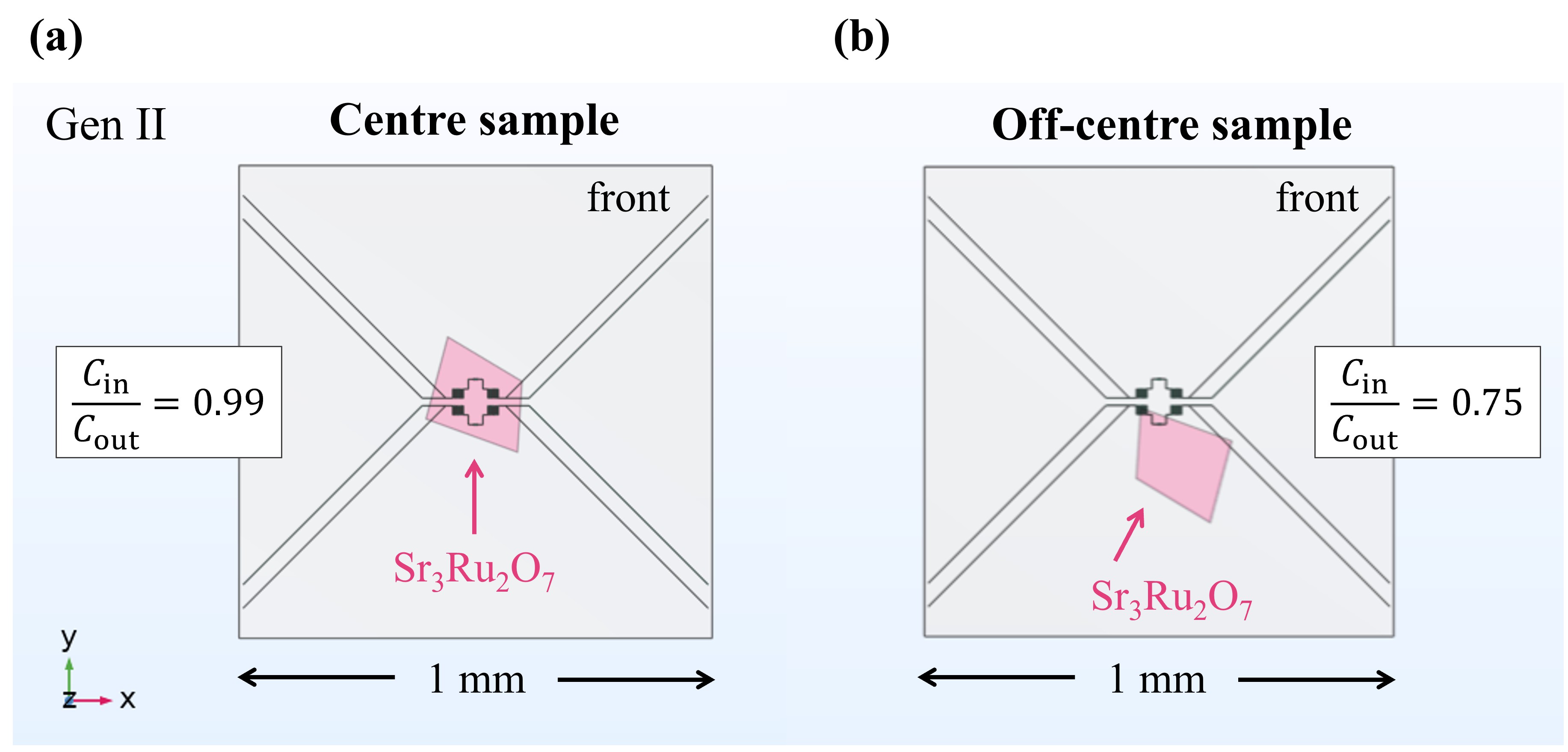}
    \caption{The calorimeter geometry, with two sample placement.
    The factor \(C_{in}/C_{out}\) is the correction factor calculated using COMSOL.
    The ideal case is shown in \textbf{a}: The sample is covering both the heater and thermometer, the measured heat capacity in this case only has a 1~\si{\percent} error.
    \textbf{b}: In the non-ideal case, the sample is only covering the thermometer.
    With this arrangement, there is a much larger systematic error on the measured heat capacity of 0.75.}
    \label{fig:off center correction}
\end{figure}

\begin{figure}[h]
    \centering
    \includegraphics[width=0.75\linewidth]{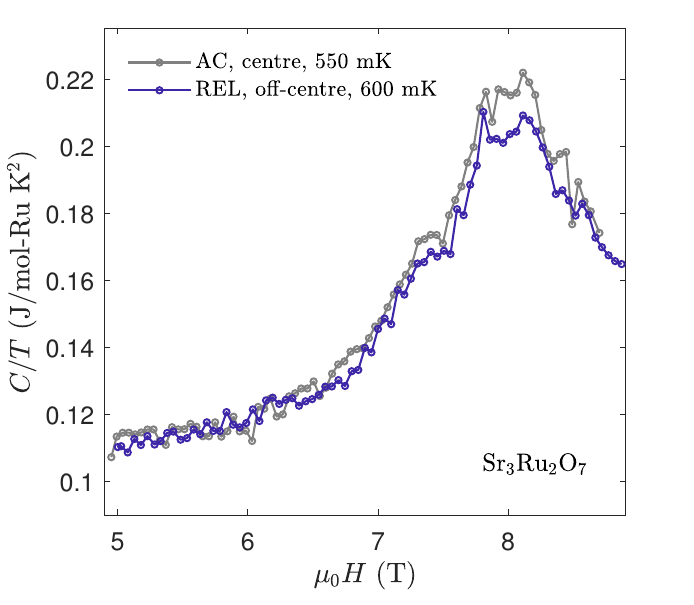}
    \caption{A comparison of the relaxation method measured $C/T$ for an off-center sample placement, with the AC method curve for a well placed one. 
    The high degree of agreement between the two shows the numerically calculated correction factor for sample position to be accurate.
    }
    \label{fig:327 specific heat offcenter comparison}
\end{figure}

We have performed measurements with an off center sample in order to validate these simulations.
The same microcrystal of \ce{Sr3Ru2O7} was used as for the above measurements, however now positioned so as to only cover one of the CBTs as shown in Fig.~\ref{fig:off center correction}.
Fig.~\ref{fig:327 specific heat offcenter comparison} shows corrected specific heat from DC measurements on top of ideal placement AC data from above.
The two agree almost exactly, highlighting the power of the numerical simulations in accounting for less than ideal experimental realities, greatly increasing the flexibility of the setup.

\section{\label{ap:transition extraction}Extraction of transition widths from \ce{CeRh2As2} dataset.}

\begin{figure}
    \centering
    \begin{overpic}[width=0.75\linewidth]{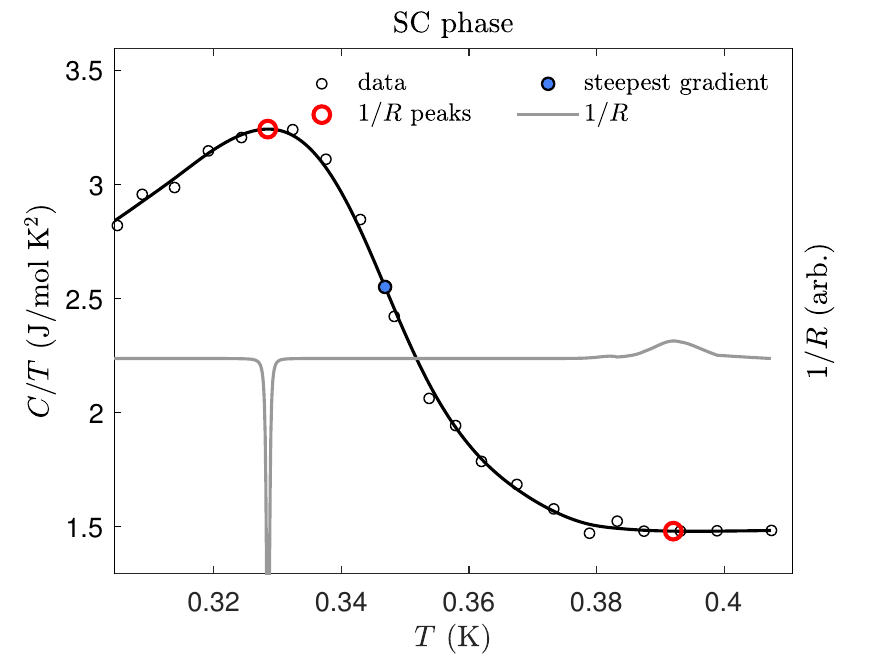}
        \put (0, 70) {(a)}
    \end{overpic}
    
    \begin{overpic}[width=0.75\linewidth]{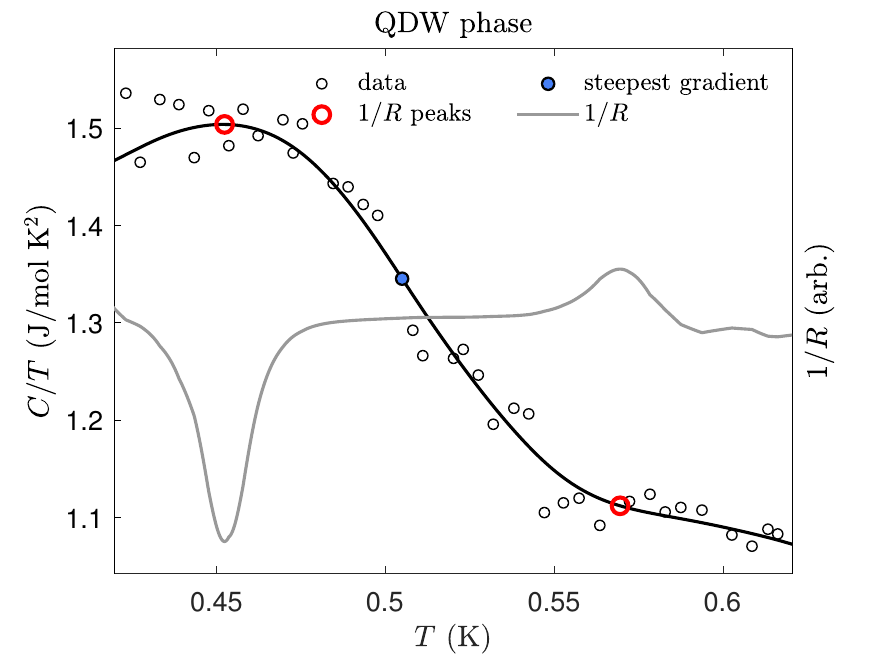}
        \put (0, 70) {(b)}
    \end{overpic}
    
    \caption{Figures showing how the transition temperatures are extracted for \textbf{a} the SC phase, and \textbf{b} the QDW phase, for $C/T$ at 1~\si{\tesla}. The left axis is the measured $C/T$, with the data shown as points and the solid black line a smoothing spline. The transition midpoint is marked on both figures and defined as the point of steepest gradient. The minimum and maximum transition temperatures are defined according to points of minimum curvature $R$. To identify these, the inverse of curvature $1/R$ is plotted on the right axis. The minimum and maximum peaks of $1/R$ correspond to the minimum and maximum transition temperature respectively.}
    \label{ap:fig:transtion extraction}
\end{figure}

As the density wave transition is weak and very broad, especially at low temperature, a systematic method of extracting the transition onset and completion was required.
In the past an equal entropy construction has been used for this,\cite{Semeniuk2023} however this does not always line up with the obvious kinks in the traces.
For this reason we use local definitions based on the derivatives of splines fit to the experimental data.
The point of fastest rate of change is used for the stated transition temperatures.
For the transition widths we focus on kinks in the data, which correspond to points of small radius of curvature $R$ defined as
\begin{equation}
    R = \frac{\left(1 + y'^2\right)^{3/2}}{y''}\;,
\end{equation}
where y is $C/T$ and $'$ denotes the temperature or field derivative depending on the experiment.
To find these areas of low curvature peaks in $1/R$ are used.
The onset of the transition appears as a transient increase in $1/R$ in an area of locally positive curvature while the completion is a sharp peak in a locally negative area.
This is shown in Fig.~\ref{ap:fig:transtion extraction}.
The transition width is then taken as the temperature/field between these endpoints.

\section{\label{ap:data aquisition and processing}Details of data processing}

\subsection{Relaxation Measurements}
\begin{figure}
    \centering
    \includegraphics[width=0.95\linewidth]{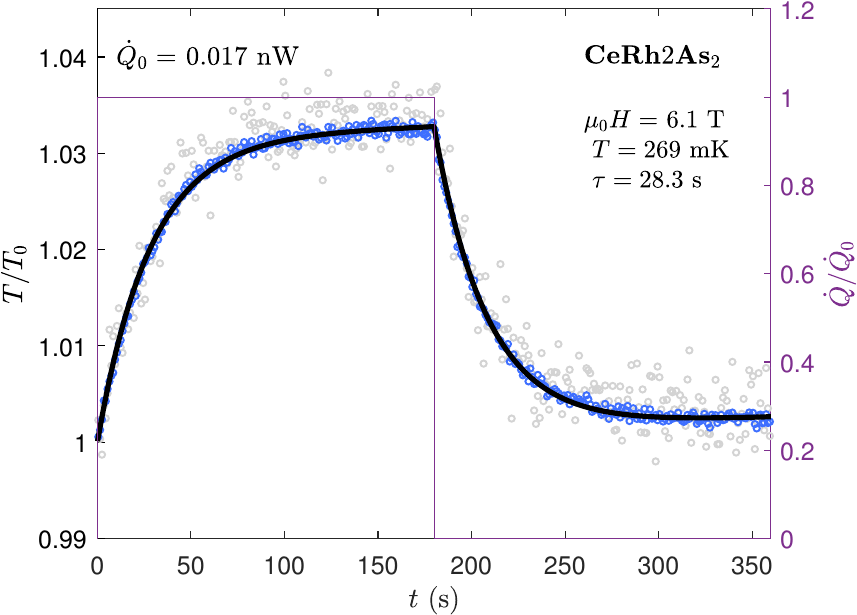}
    \caption{An example time trace from a DC relaxation measurement. The points are binned data from 10 consecutive heat pulses. The solid black line is a fit to the standard relaxation equation (\ref{eqn:DC responce}) used to extract \(\Delta T\)\ and \(\tau\). The purple line shows the applied power (right axis).}
    \label{ap:fig:relaxation measurment}
\end{figure}

Relaxation measurements were performed with the standard heat pulse method\cite{Bachmann1972} using the SPECS Tramea measurement system to both drive the heater as well as record \(G_0\), heater current and voltage utilizing the Tramea's real time FPGA with \qty{10}{\micro\second} time accuracy allowing completely synchronized, repeatable measurements.
The timing accuracy enables for example simple measurement cycle averaging with no loss of accuracy in order to improve signal to noise.
\(G_0\) was converted to temperature using the calibration and equations in section \ref{sec:CBT background}, with subsequent analysis of \(T\) using equation (\ref{eqn:DC responce}).
Both heating and cooling were fit simultaneously forcing the same \(\Delta T\) and \(\tau\). (The fit when appropriate (during for example slow field sweeps) also included a linear drift term).
An example heat-cool cycle along with a ten-cycle average is shown in Fig.~\ref{ap:fig:relaxation measurment}.
We established that either fitting individual pulses and averaging \(\tau\) or fitting averaged pulses to extract a single \(\tau\) gives the same result within the uncertainty. (In the  example shown \(\left(\tau=28.3\pm0.3\right)\)~\si{\second} and \(\left(\tau=28.4\pm0.3\right)\)~\si{\second} respectively.)

\subsection{AC measurements}
\begin{figure}
    \centering
    \includegraphics[width=0.95\linewidth]{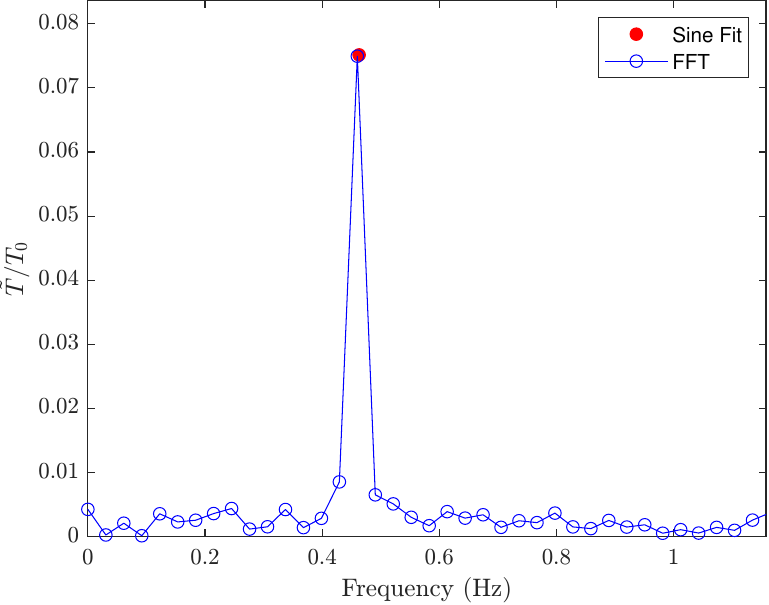}
    \caption{An FFT of the \qty{658}{\milli\kelvin} data as well as the results of the fit presented in the inset of Fig~\ref{fig:sr327-freqsweep} showing good agreement between the two processing techniques and the possibility of operating the nanocalorimeter using a lock in amplifier.}
    \label{ap:fig:AC_FFT}
\end{figure}

AC measurements likewise utilized the Tramea measurement system to apply an AC excitation to the heater.
After conversion from \(G_0\) to \(T\), data processing could proceed in one of two ways, either using a sine fit or via Fourier analysis (essentially a software Lock-in amplifier analysis).
We found fitting a sine wave to the temperature using short batches of roughly 10 oscillations the simplest process.
An example of this is shown in the inset in Fig.~\ref{fig:sr327-freqsweep}.
We have found Fourier analysis to give the same result, with a Fourier transform of the data from Fig.~\ref{fig:sr327-freqsweep} shown along with the results of the sine fit in Fig.~\ref{ap:fig:AC_FFT}.
Once extracted, the amplitude of the temperature oscillation was converted to heat capacity as discussed in the main text.

\section{Thermal Gradients in the sample}

\begin{figure}
    \centering
    \includegraphics[width=0.95\linewidth]{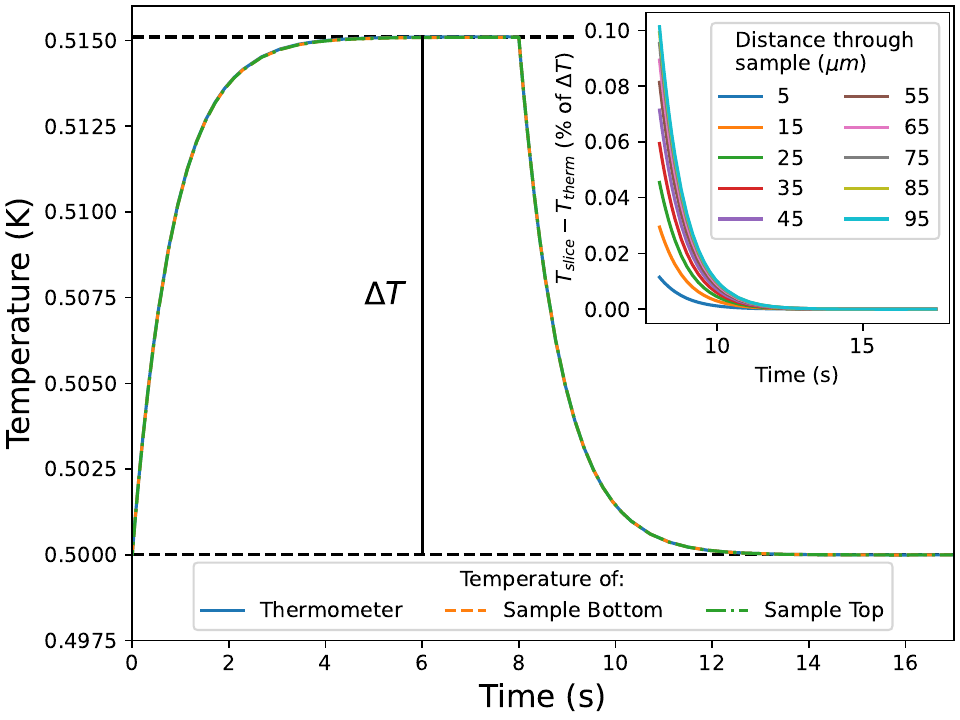}
    \caption{Simulated thermal gradients in the sample along \(z\) (the membrane normal) during a DC relaxation measurement of \ce{Sr3Ru2O7}. Main: Temperature of the thermometer as well as the top and bottom \qty{10}{\micro\meter} of the sample as a function of time (the bottom of the sample is in contact with the membrane). The heater is turned on at \qty{0}{\second} and off at \qty{8}{\second}. Inset: The difference between each \qty{10}{\micro\meter} slice of the sample and the thermometer during the cool down showing the sample equilibrating. The maximum thermal gradient is \qty{0.1}{\percent} of \(\Delta T\).}
    \label{ap:fig:327_thermal_gradients}
\end{figure}

The FEM simulations can also be used to confirm the lack of any major thermal gradients within the sample along the \(z\) direction (normal of the membrane).
To estimate this effect we simulated a \qty{500}{\milli\kelvin} bath temperature measurement using a sample modeled on the actual \ce{Sr3Ru2O7} sample used in section \ref{sec:validation}.
Fig.~\ref{ap:fig:327_thermal_gradients} shows the time evolution of the temperature as 'measured' by the simulated thermometer CBT on the membrane as well as the average temperature in the first and last \qty{10}{\micro\meter} of sample along the direction normal to the membrane.
As the difference is negligible compared to the size of the heat pulse \(\Delta T\) and therefore not visible in this figure we show in the inset the process of equilibration during the cool down period, plotted as the difference between the average temperature in each \qty{10}{\micro\meter} slice of sample and the thermometer. 
The maximum thermal gradient across the sample is extremely small at just \qty{0.1}{\percent} of \(\Delta T\) or \qty{1.5}{\micro\kelvin}, occurring immediately after the heater turns off and dropping rapidly as the sample equilibrates.

\bibliography{bib}% Produces the bibliography via BibTeX.

\end{document}